\documentclass[lettersize,journal]{IEEEtran}
\usepackage{amsmath,amsfonts}
\usepackage{algorithmic}
\usepackage{algorithm}
\usepackage{array}
\usepackage[caption=false,font=footnotesize,labelfont=rm,textfont=rm]{subfig}
\usepackage{textcomp}
\usepackage{stfloats}
\usepackage{url}
\usepackage{verbatim}
\usepackage{graphicx}
\usepackage{subfig}
\usepackage{cite}
\usepackage{amsmath,array}
\usepackage{bm}
\begin{document}

\title{Toward Robust Semantic Communications: Proactive Importance-Ordered Restructuring for Enhanced Unequal Error Protection}

\author{Xunyang Zhan,~\IEEEmembership{Student Member,~IEEE,}
Jie Cao,~\IEEEmembership{Member,~IEEE,}
Xu Zhu,~\IEEEmembership{Senior Member,~IEEE,} \\
Nikolaos Pappas,~\IEEEmembership{Senior Member,~IEEE,}
Zhijin Qin,~\IEEEmembership{Senior Member,~IEEE,}
and~Shaohan Feng,~\IEEEmembership{Member,~IEEE}
\thanks{Corresponding author: Jie Cao.}
\thanks{Xunyang Zhan is with the School of Information Science and Technology, Harbin Institute of Technology, Shenzhen 518055, China (e-mail: 24s052039@stu.hit.edu.cn).}
\thanks{Jie Cao and Xu Zhu are with the School of Information Science and Technology, Harbin Institute of Technology, Shenzhen 518055, China, also with the Guangdong Provincial Key Laboratory of Aerospace Communication and Networking Technology, Shenzhen 518055, China, and also with the Shenzhen Municipal Key Laboratory of AIoT Communications, Shenzhen 518055, China (e-mail: caojhitsz@ieee.org; xuzhu@ieee.org).}
\thanks{Nikolaos Pappas is with the Department of Computer and Information Science, Linköping University, 58183 Linköping, Sweden (e-mail: nikolaos.pappas@liu.se).}
\thanks{Zhijin Qin is with the Department of Electronic Engineering, Tsinghua University, Beijing 100084, China, also with the State Key Laboratory of Space Network and Communications, Beijing 100084, China, and also with Beijing National Research Center for Information Science and Technology, Beijing 100084, China (e-mail: qinzhijin@tsinghua.edu.cn).}
\thanks{Shaohan Feng is with the School of Information and Electronic Engineering, Zhejiang Gongshang University, Hangzhou 310018, China (e-mail: feng\_shaohan@mail.zjgsu.edu.cn).}
}




\maketitle

\begin{abstract}
Semantic communications (SemCom) is a promising task-oriented paradigm in which semantic features exhibit non-uniform importance. Consequently, unequal error protection (UEP), which allocates resources based on semantic importance, plays a pivotal role in maximizing system utility.
However, most existing schemes adopt passive importance evaluation, which neither proactively reshapes the importance distribution nor explores its impact on UEP performance. 
In this paper, we propose a novel importance-ordered semantic feature restructuring (ISFR) scheme that proactively enforces a descending importance hierarchy and jointly optimizes multi-dimensional resources to improve system utility.
Specifically, modules with decreasing retention probabilities and increasing distortion levels are employed, which drive the model to concentrate key semantics into front-end features and thus strengthen importance differentiation.
Moreover, a joint optimization problem that jointly optimizes channel matching, feature selection, modulation schemes, and power allocation is formulated to minimize the importance-weighted total semantic distortion. 
To solve this non-convex problem, a hierarchical decoupling strategy is proposed, which decomposes it into four tractable subproblems.
This approach leverages the ordered prior to drastically prune the search space for feature selection and modulation, while integrating greedy-based channel matching and convex power allocation.
Simulation results demonstrate that the proposed ISFR scheme outperforms traditional uniform importance-based schemes under harsh channel conditions and limited resources, validating the significant robustness improvement enabled by the concentration of key semantic information. 
\end{abstract}


\newpage
\section{Introduction}
\IEEEPARstart{T}{he} evolution of sixth-generation (6G) wireless networks has spurred diverse intelligent applications, such as virtual reality, digital twins, and holographic communications\cite{survey-6G}. These applications generate massive data, imposing stringent requirements on system capacity, latency, and reliability. However, traditional lossless communication paradigms are increasingly challenged by explosive data growth under limited spectrum and energy resources\cite{survey-SC1}. To overcome these bottlenecks, semantic communications (SemCom) has emerged as a transformative paradigm. By extracting and transmitting only task-relevant information, SemCom significantly reduces data traffic and enhances system efficiency, offering a promising solution for future intelligent networks\cite{deepsc,dong1,survey-SC2,semcom10}. 

Unlike traditional systems that emphasize the transmission process and treat all data indiscriminately, SemCom is inherently task-oriented\cite{cao1}. Specifically, traditional systems evaluate performance using bit error rate (BER) and throughput, aiming to minimize transmission errors across all transmitted bits. In contrast, SemCom targets maximizing task performance, where semantic features contribute unequally to downstream tasks\cite{cao2}. Moreover, wireless environments are heterogeneous, with fluctuating channel states, varying resource availability, and diverse user requirements. Consequently, enforcing uniform error performance across all features leads to inefficient resource utilization and suboptimal task performance, especially in resource-constrained scenarios.

The dual heterogeneity in semantic importance and transmission environments necessitates aligning communication resource allocation with the semantic contribution of transmitted data. 
Inspired by classical unequal error protection (UEP) in information theory, which provides differentiated reliability at the bit or packet level via coding and modulation, we extend this concept to SemCom as semantic importance-aware UEP (SI-UEP). SI-UEP characterizes the task contribution of semantic features as semantic importance, then leverages it to guide differentiated transmission and resource allocation, thereby achieving non-uniform reliability. Under harsh channels and limited resources, SI-UEP prioritizes critical semantics to enhance task robustness against channel impairments.

\subsection{Related Work}
\textit{SemCom:} The rapid development of deep learning has driven breakthroughs in SemCom. In \cite{DJSCC}, a classic deep joint source-channel coding (DJSCC) framework was introduced. Technologies including attention mechanism\cite{ADJSCC}, variable-length coding\cite{DJSCC-V}, rate-distortion optimization\cite{NTSCC}, and diffusion model\cite{Diffusion} have been incorporated to further enhance SemCom systems. Moreover, integration with existing digital communication systems has been realized via learnable quantization \cite{LSQ} and joint coding-modulation \cite{JCM}, while flexible retransmission strategies have been applied\cite{HARQ}. Multi-user\cite{SC-MU}, multi-antenna\cite{SC-MIMO} and multi-access\cite{SC-NOMA} scenarios have been investigated to improve system capacity. In addition, the orthogonality of semantic features\cite{SC-SFDMA} and models\cite{SC-OMDMA} has been explored to mitigate interference. Meanwhile, adaptive resource allocation strategies have been designed to optimize system efficiency\cite{Resource1}. However, these schemes fail to explore SI-UEP strategies based on the inherent importance differences among semantic information.

\textit{Modeling of Semantic Importance:} Semantic importance prioritizes features according to source-channel-task dependencies. Existing approaches include gradient analysis\cite{Grad-CAM,grad1,grad2,grad-corr,grad3}, robustness verification\cite{masking,grad-act,rubust1,rubust2}, pre-trained model evaluation\cite{atten2,com4cv,LLM}, learning-based partitioning\cite{eval-net,learned-BER1,learned-BER2}, and information timeliness modeling\cite{aoi1,aosi,uoi,aoi2}. 
Gradient-weighted class activation mapping (Grad-CAM) \cite{Grad-CAM} is a representative method that quantifies feature contributions using average gradients \cite{grad1,grad2,grad-corr}. Moreover, knowledge distillation has been employed in \cite{grad3} to address task performance unavailability at the transmitter.
Robustness verification quantifies importance by performance sensitivity to feature perturbations, including feature-wise masking \cite{masking} and its gradient-activation approximation \cite{grad-act}, as well as decoder-side linear relaxation \cite{rubust1,rubust2}.
In contrast, pre-trained model evaluation leverages general model parameters, such as attention scores in vision Transformers \cite{atten2}, segmentation saliency maps \cite{com4cv}, and embedding cosine similarity \cite{LLM}. Despite simple implementation, these methods often lack task specificity.
Beyond explicit modeling, importance can also be implicitly learned via end-to-end training, either by jointly training an evaluation network \cite{eval-net} or treating BER as trainable parameters to characterize bit-level importance \cite{learned-BER1,learned-BER2}.
Regarding timeliness, age of information (AoI) \cite{cao-aoi} captures temporal priority but ignores semantics, while semantic-aware metrics such as semantics of information\cite{aoi1}, age of incorrect semantics \cite{aosi}, and utility loss of information \cite{uoi} balance timeliness, reliability, and overhead \cite{aoi2}.
However, most existing methods adopt a post-training evaluation paradigm, yielding importance weights inherently tied to specific model architectures. Crucially, they overlook strategies to proactively reshape the importance distribution for more efficient transmission.

\textit{Applications of Semantic Importance:} Aligning transmission mechanisms with semantic importance is essential for optimizing SemCom systems, particularly under heterogeneous channel and resource conditions. Representative applications include feature selection, feature scheduling, rate control, and resource allocation. In resource-constrained scenarios, feature selection optimizes compression ratios to prioritize high-importance feature transmission \cite{grad1,grad2,masking,grad-act}. Conversely, to mitigate severe channel variations, feature scheduling assigns critical features to reliable physical resources \cite{grad3,eval-net}. In addition, adaptive rate control enhances the rate-distortion trade-off by dynamically adjusting parameters such as the number of semantic symbols, quantization bits, and coding rates \cite{rubust1,rubust2,LLM,aoi1}. Furthermore, incorporating semantic importance into multi-domain resource optimization (\textit{e.g.} time, frequency, space, and power) enables efficient transmission with reduced overhead \cite{grad-corr,learned-BER1,learned-BER2,grad1,atten2,com4cv,aosi,uoi}.
However, most existing strategies allocate resources based on pre-determined weights. Such inherently learned importance distributions often exhibit limited differentiation, creating a bottleneck for SI-UEP. Consequently, exploring distribution characteristics conducive to SI-UEP remains an open issue.

\subsection{Contributions}
Motivated by the open issues, we investigate the proactive importance-ordered feature restructuring and its interplay with SI-UEP strategies in wireless SemCom systems to answer the following questions:
\begin{itemize}
    \item[1)] How does feature importance differentiation affect SI-UEP efficacy under harsh channels and limited resources?
    \item[2)] How can we proactively reshape the feature importance distribution during model training?
	\item[3)] How can we leverage these importance characteristics to design an efficient SI-UEP resource allocation strategy?
\end{itemize}

The main contributions of this paper are summarized as follows.
\begin{itemize}
\item To the best of our knowledge, this is the first work to investigate the impact of feature importance differentiation on SI-UEP efficacy in SemCom systems. A novel importance-ordered semantic feature restructuring (ISFR) scheme is proposed to proactively induce importance differentiation, upon which multi-dimensional resource allocation is optimized to maximize system utility. Compared to traditional schemes that exhibit near-uniform importance (UI) \cite{grad1,grad2,grad3,grad-corr,masking,grad-act}, ISFR achieves significant performance gains, particularly under adverse conditions with low signal-to-noise ratio (SNR), limited power, and high rate. While this introduces a trade-off between robustness and informativeness, it involves a strategic sacrifice of marginal 1.3\% peak performance in exchange for over 23\% improvement in robustness. Consequently, significant importance differentiation is fundamentally superior to uniformity for SI-UEP, as it ensures limited resources are concentrated on protecting the most critical semantics.
\item A robust training strategy is proposed to realize ISFR, ensuring a descending hierarchy of feature contributions. Specifically, a BER matching module and a nested dropout module are designed to impose decreasing retention probabilities and increasing distortion levels across features. These mechanisms induce the model to concentrate critical semantics into front-end features autonomously, marking a paradigm shift from passive importance evaluation in previous works\cite{grad1,grad2,grad3,grad-corr,masking,grad-act} to proactive restructuring. It constructs a structured, hierarchical semantic representation, unlike implicit channel-adaptive schemes based on attention mechanisms \cite{ADJSCC}. Moreover, this strategy is low-complexity, independent of inherent data saliency (\textit{e.g.}, spatial or modal characteristics \cite{com4cv,rubust1}), and incorporates randomness to enhance generalization across diverse scenarios. Simulation results confirm that the importance-weight differentiation across features can span up to two orders of magnitude. 
\item To minimize the weighted total semantic distortion, an SI-UEP problem is formulated, through joint optimization of feature selection, channel matching, modulation schemes, and power allocation. This formulation offers a more comprehensive framework than existing SI-UEP schemes \cite{grad3,com4cv,learned-BER1}. To address the intractable problem, an ordered prior-assisted hierarchical decoupling (OPHD) algorithm is proposed. Specifically, separable channel matching is first solved using a greedy approach, while the remaining problem is transformed into a nested iterative search. By exploiting the ordered importance prior, the search space for features and modulation is significantly pruned, reducing the complexity from exponential to linear. Finally, power allocation is formulated as a convex optimization problem. The OPHD algorithm delivers a low-complexity and near-optimal solution, validating the advantage of ISFR for efficient resource optimization. Moreover, this work highlights the necessity of integrating semantic importance as an indispensable dimension in UEP.
\end{itemize}


\begin{figure*}[!t]
\centering
\includegraphics[width=16cm]{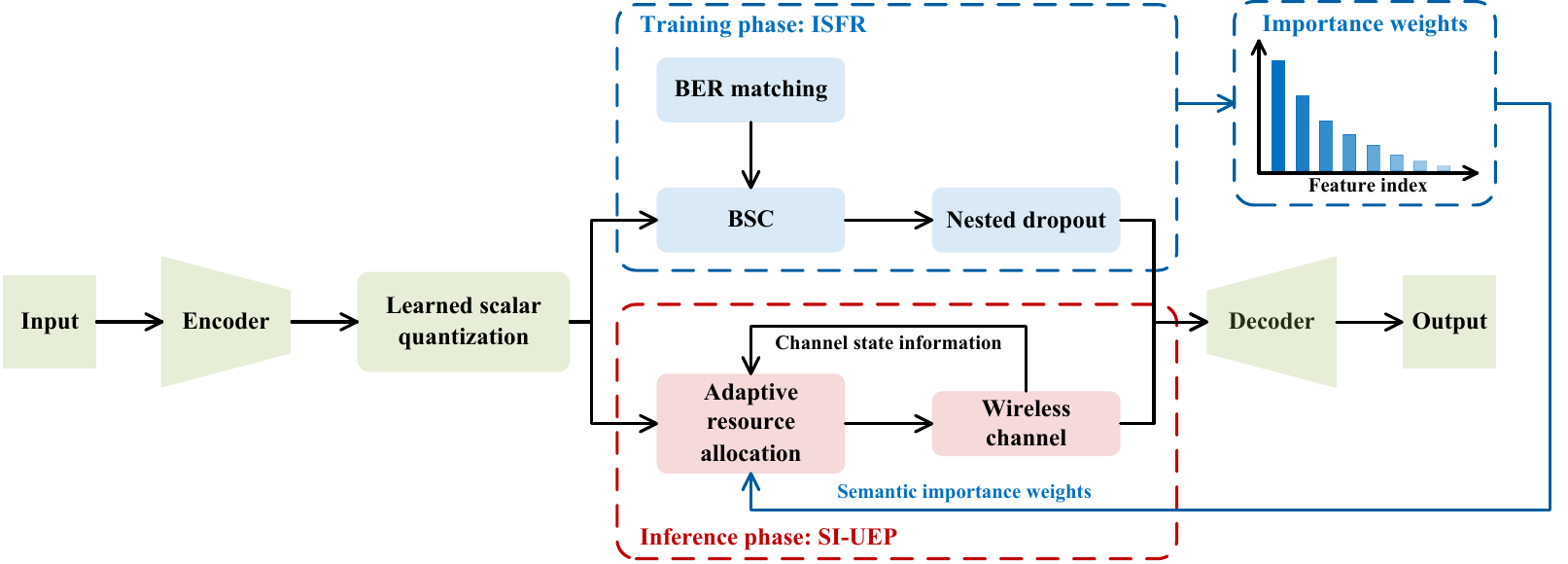}
\caption{System model of the proposed ISFR and SI-UEP schemes. During training, ISFR induces a feature distribution with descending semantic importance via the BER matching and nested dropout modules, and the corresponding importance weights are evaluated and stored as prior. During inference, SI-UEP is performed via multi-dimensional resource allocation driven by importance and CSI.}
\label{System_model}
\vspace{-0.2cm}
\end{figure*}

\section{System Model}
In this section, we present the system model of the proposed ISFR and SI-UEP schemes.

\subsection{Overall Framework}
As illustrated in Fig. \ref{System_model}, we consider a digital SemCom system for image transmission consisting of two tightly coupled phases: the training phase for ISFR and the inference phase for SI-UEP. During training, ISFR restructures features to exhibit descending semantic importance via the BER-matching and nested dropout modules. The resulting importance weights are then evaluated and stored as prior knowledge. During inference, multi-dimensional resource allocation is jointly optimized based on the importance prior and real-time channel state information (CSI), enabling efficient SI-UEP to minimize total semantic distortion. In this way, the former phase establishes an ordered semantic representation offline, while the latter exploits this prior for channel-adaptive transmission online.

From the overall signal flow, the source image is first encoded into semantic features, then quantized and converted into bitstream for digital transmission, and finally reconstructed at the receiver. Specifically, the transmitter encodes the original image $\mathbf{X}$ using a $\theta$-parameterized encoder $ f(\cdot;\theta)$ to generate the continuous feature $\mathbf{Y}$, given by
\begin{equation}
\mathbf{Y} = f(\mathbf{X};\theta) \in \mathbb{R}^{L \times N},\label{eq1}
\end{equation}
where $L$ and $N$ denote the dimension and number of features, respectively, through grouping features in a specific manner (\textit{e.g.}, channel-wise in convolutional neural networks (CNN)). Then, each element of $\mathbf{Y}$ is quantized into $\mathbf{Z}$ to obtain a discretized semantic representation, expressed as
\begin{equation}
z_{i,j} = g_\text{q}(y_{i,j}) \in \mathbf{Q}, \ i \in \{1, \cdots, L\}, j \in \{1, \cdots, N\},\label{eq2}
\end{equation}
where $y_{i,j},z_{i,j}$ denote the $i,j$-th elements of $\mathbf{Y}$, $\mathbf{Z}$, and $\mathbf{Q} = \{ \tilde{q}_1, \tilde{q}_2, \cdots, \tilde{q}_{2^{N_\text{b}}} \}$ represents the quantization levels achievable with $N_\text{b}$ bits. Since the output of semantic encoder approximately follows a normal distribution, the learnable scalar quantization proposed in \cite{LSQ} is adopted to reduce quantization error. The quantization function is defined as
\begin{equation}
g_\text{q}(y_{i,j}) = \Delta_0 + \sum_{t=1}^{2^{N_\text{b}} - 1} (\Delta_t - \Delta_{t-1}) \cdot \sigma(y_{i,j}-\Delta_t; T),\label{eq3}
\end{equation}
where $\Delta_t$ is the $t$-th level in the learnable quantization set, $T$ is the temperature coefficient controlling the annealing of the approximation function from soft to hard, and $\sigma(\cdot)$ is the Sigmoid function defined as $\sigma(x; T) = \frac{1}{1 + \exp(-Tx)}$.
The discrete quantization levels $\mathbf{Z}$ are converted into a bitstream $\mathbf{B}$, which is given by
\begin{equation}
\mathbf{B} = g_\text{ad}(\mathbf{Z}) \in \{0, 1 \}^ {(L \times {N_\text{b})} \times N},\label{eq4}
\end{equation}
where $g_\text{ad}(\cdot)$ maps quantization levels to bits. At the receiver, the image $\mathbf{\hat{X}}$ is reconstructed from the distorted feature $\hat{\mathbf{Z}}$ via a $\phi$-parameterized decoder $ h(\cdot;\phi)$, given by $\mathbf{\hat{X}} = h(\hat{\mathbf{Z}};\phi)$.

The encoder, quantization, bit mapping, and decoder constitute the backbone of the proposed system and operate in both phases. Built upon this, the training and inference phases process the bitstream $\mathbf{B}$ differently, corresponding to the two branches in Fig. \ref{System_model}. These phase-specific designs are introduced in the following two subsections.

\subsection{Training Phase for ISFR}
In the training phase, the bitstream $\mathbf{B}$ is further processed to induce a feature distribution with decreasing importance across feature indices. Specifically, the semantic bitstream sequentially passes through a binary symmetric channel (BSC) with the BER matching module, a bit to quantization level mapping module, and the nested dropout module. Their detailed design will be discussed in Section III. This process is expressed as
\begin{equation}
\hat{\mathbf{Z}} = g_\text{nd}(g_\text{da}(g_\text{bm}(\mathbf{B}))),\label{eq5}
\end{equation}
where $g_\text{bm}(\cdot)$ and $g_\text{nd}(\cdot)$ denote the BER-matched BSC and the nested dropout functions, respectively, and $g_\text{da}(\cdot)$ is the bit to quantization level mapping. 

The parameters $\{ \theta,\phi,\Delta \}$ of the encoder, decoder, and quantization module are jointly trained to realize ISFR. After training, the semantic importance weights of each feature are evaluated on the validation dataset $\mathbf{X}_\mathrm{val}$ using the function $g_\text{w}(\cdot)$ and stored as prior knowledge, which is expressed as
\begin{equation}
\mathbf{w} = g_\text{w}(\mathbf{X}_\mathrm{val};\theta;\phi;\Delta) \in \mathbb{R}^{N}.\label{eq6}
\end{equation}

\begin{figure*}[!t]
\setlength{\abovecaptionskip}{-0.05cm}   
\centering
\includegraphics[width=14cm]{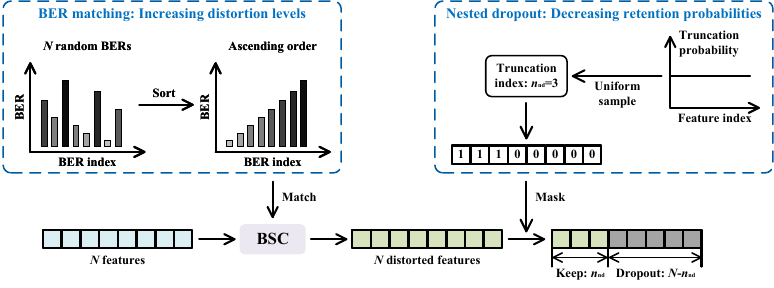}
\caption{System model of the BER matching module and nested dropout module for ISFR.}
\label{ISFR}
\vspace{-0.1cm}
\end{figure*}

\subsection{Inference Phase for SI-UEP}
During inference, the importance weights are adopted as a prior to guide SI-UEP. In detail, the bitstream $\mathbf{B}_k$ of $k$ retained features is first converted into channel symbols $\mathbf{S}$ using $2^{\mathbf{m}}$ order quadrature amplitude modulation (QAM) and allocated transmission power $\mathbf{p}$. 

The objective of SI-UEP is to enhance task performance by minimizing the importance-weighted total semantic distortion. Accordingly, the transmitter jointly optimizes channel matching, feature selection, modulation schemes, and power allocation based on the prior importance weights $\mathbf{w}$, CSI $\boldsymbol{\gamma}$, power budget $P_{\text{max}}$, and rate budget $M_{\text{min}}$, given by
\begin{equation}
\{ k^*, \boldsymbol{\gamma}^*, \mathbf{p}^*, \mathbf{m}^* \} = g_\text{ra}(\mathbf{w},\boldsymbol{\gamma}, P_{\text{max}}, M_{\text{min}}).\label{eq7}
\end{equation}
By discarding low-importance features and reallocating superior resources to high-importance ones (\textit{e.g.}, more reliable channel, higher power, and lower modulation order), SI-UEP significantly reduces the distortion of task-critical features, thereby enhancing overall task performance. 

The wireless channel is modeled as a block fading channel, in which the channel coefficients remain constant over the coherence time. Under this assumption, each received symbol $\mathbf{\hat{s}}_j$ and the corresponding SNR are expressed as
\begin{equation}
\hat{\mathbf{s}}_j = \sqrt{p_j}h_j\mathbf{s}_j + \mathbf{n}_j,\label{eq8}
\end{equation}
\begin{equation}
\text{SNR}_{j}=\frac{p_j|h_j|^{2}}{\sigma^2}=p_j\gamma_j,\label{eq9}
\end{equation}
where $h_j \in \mathbb{C}$ is the complex-valued channel coefficient, $n_j\sim\mathcal{CN}(0,\sigma^2)$ denotes additive white Gaussian noise (AWGN), $p_{j}$ is the power of each symbol in $\mathbf{s}_j$, and $\gamma_j=\frac{|h_j|^{2}}{\sigma^2}$ is the normalized SNR (\textit{i.e.}, SNR with unit transmit power). Assuming perfect channel estimation, the equalized symbol is given by $\tilde{\mathbf{s}}_j=\frac{h_j^*}{|h_j|^2}\mathbf{\hat{s}_j}$,
where $h_j^*$ is the complex conjugate of $h_j$. The received feature $\hat{\mathbf{Z}}$ is then obtained from $\tilde{\mathbf{S}}$ through demodulation $g_\text{dm}(\cdot)$ and inverse mapping $g_\text{da}(\cdot)$, given by
\begin{equation}
\hat{\mathbf{Z}} = g_\text{da}(g_\text{dm}(\tilde{\mathbf{S}})).\label{eq10}
\end{equation}

\section{Module Design and Training Strategy \\ for ISFR}
In this section, we introduce a BER matching module and a nested dropout module, as illustrated in Fig. \ref{ISFR}. By leveraging unequal retention probabilities and distortion levels, these modules induce the concentration of key semantics via a two-stage training strategy.

\subsection{BSC with BER Matching Module}
The BSC is adopted to model the digital transmission process. This choice allows the aggregate effects of diverse channel states and transmission parameters to be characterized by a single metric, \textit{i.e.}, the BER $\mu$. Consequently, the trained model has sufficient adaptability and generalization capabilities across different scenarios. 

Since SI-UEP relies on differentiated protection through unequal resource allocation, it necessitates significant variation in semantic importance across features. To achieve ISFR with a descending importance order, unequal perturbations are applied to features. Specifically, smaller perturbations are imposed on front-end features to cause minor distortions, while larger perturbations are imposed on tail-end features to cause significant distortions. Driven by this unequal distortions, the model is induced to concentrate key semantic information into the front-end features, thereby establishing an ordered semantic contribution.

In digital communication systems, unequal perturbations can be realized by imposing unequal BERs on feature bitstreams. Therefore, a BER matching module is designed. It generates $N$ random BERs $\textbf{BER}_{\text{bm}}$ following a uniform distribution within $[0,\text{BER}_\text{max}]$, and sorts them in ascending order. These sorted BERs are then assigned to the $N$ feature bitstreams, imposing perturbations of increasing intensity. This process is expressed as
\begin{equation}
\textbf{BER}_{\text{bm}} \sim \text{Uniform}(0, \text{BER}_{\text{max}}) \in \mathbb{R}^{N \times 1},\label{eq15}
\end{equation}
\begin{equation}
\mathbf{B}_\text{bm} = g_\text{bm}\big( \mathbf{B};g_\text{sort}(\textbf{BER}_{\text{bm}}, \text{ascending}) \big),\label{eq16}
\end{equation}
where $g_\text{sort}(\cdot)$ denotes the sorting function. 

The BER matching module not only concentrates key semantic information into front-end features via unequal distortion levels, but also enhances model generalization across diverse channel environments through random sampling.

\begin{figure}[!t]
\setlength{\abovecaptionskip}{-0.05cm}   
\centering
\includegraphics[width=6cm]{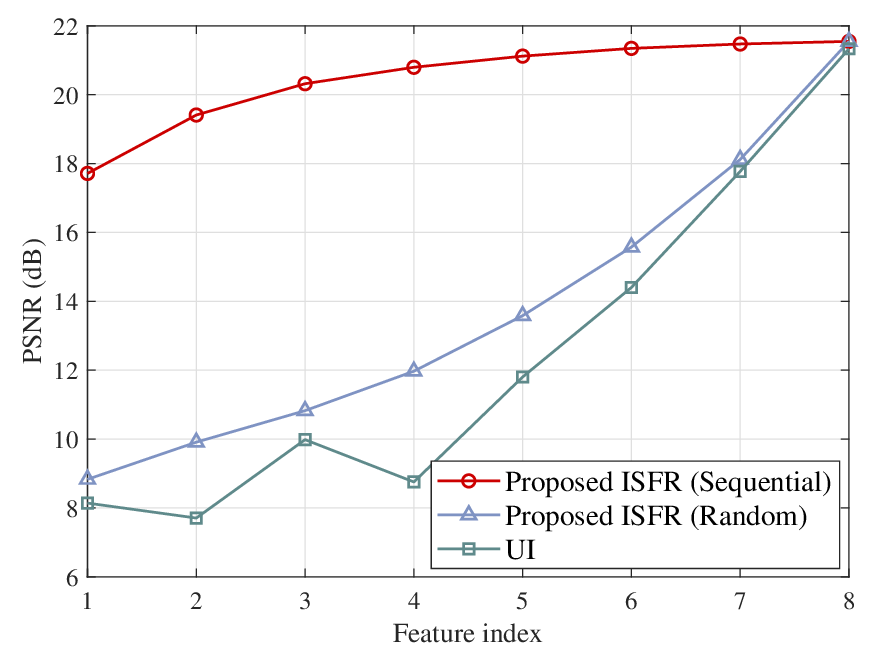}
\caption{Superposition in ISFR via feature accumulation.}
\label{imp-1}
\vspace{-0.1cm}
\end{figure}

\subsection{Nested Dropout Module}
However, relying solely on the BER matching module yields unstable ISFR performance, as demonstrated in the subsequent section. Inspired by \cite{nd-icml}, a nested dropout module is further incorporated. The core principle is to randomly truncate and discard the subsequent features. By assigning higher retention probability to front-end features and lower ones to tail-end features, the model is driven to concentrate key semantic information into the leading positions. Specifically, a truncation index $n_{\text{nd}}$ is uniformly sampled from the discrete set $\{1,2,\dots,N\}$. Based on this, a mask $\mathbf{M}_{\text{nd}}$ is generated, where the first $n_{\text{nd}}$ elements are set to $1$ and the remaining $N-n_{\text{nd}}$ elements to $0$. This mask is applied to the recovered features $\hat{\mathbf{Z}}_{\text{da}}$ to produce the decoder input, given by
\begin{equation}
n_{\text{nd}} \sim \text{Uniform}(\{1,2,\dots,N\}),\label{eq17}
\end{equation}
\begin{equation}
\mathbf{M}_{\text{nd}} = [ \underbrace{1,\ 1,\ \dots,\ 1}_{n_{\text{nd}}}, \underbrace{0,\ 0,\ \dots,\ 0}_{N-n_{\text{nd}}} ],\label{eq18}
\end{equation}
\begin{equation}
\hat{\mathbf{Z}} = \hat{\mathbf{Z}}_{\text{da}} \odot \mathbf{M}_{\text{nd}},\label{eq19}
\end{equation}

The nested structure establishes a dependency where subsequent features serve as supplements and refinements to preceding ones. Fundamentally, this scheme functions as an incremental form of progressive transmission, offering inherent advantages for variable-length coding, flexible retransmission, and SI-UEP. Furthermore, as the retention probability of the $i$-th feature is $(N+1-i)/N$, front-end features are retained and trained more frequently than tail-end ones. This training disparity enables front-end features to carry most critical semantic information, thereby enhancing the efficacy of ISFR. 

\subsection{Training Strategy for ISFR}
The simultaneous presence of learnable quantization, random BER, and random truncation indices makes stable convergence challenging for direct full model training. Consequently, a two-stage training strategy is adopted. In Stage 1, nested dropout with analog transmission is considered, as the random truncation dominates semantic representation. The objective is to train the encoder and decoder to reconstruct images using only truncated partial features. Subsequently, in Stage 2, digital transmission based on learnable scalar quantization and the BER matching module are incorporated into the training process. For image reconstruction, mean squared error (MSE) is employed as the loss function, expressed as
\begin{equation}
L(\theta, \phi, \Delta) = \| \hat{\mathbf{X}} - \mathbf{X} \| _{{\cal{L}}^2}^2.\label{eq20}
\end{equation}

This strategy, which prioritizes establishing the fundamental update direction before refining specific functions, effectively stabilizes model convergence and enhances task performance.

\begin{figure}[!t]
\setlength{\abovecaptionskip}{-0.05cm}   
\centering
\includegraphics[width=6cm]{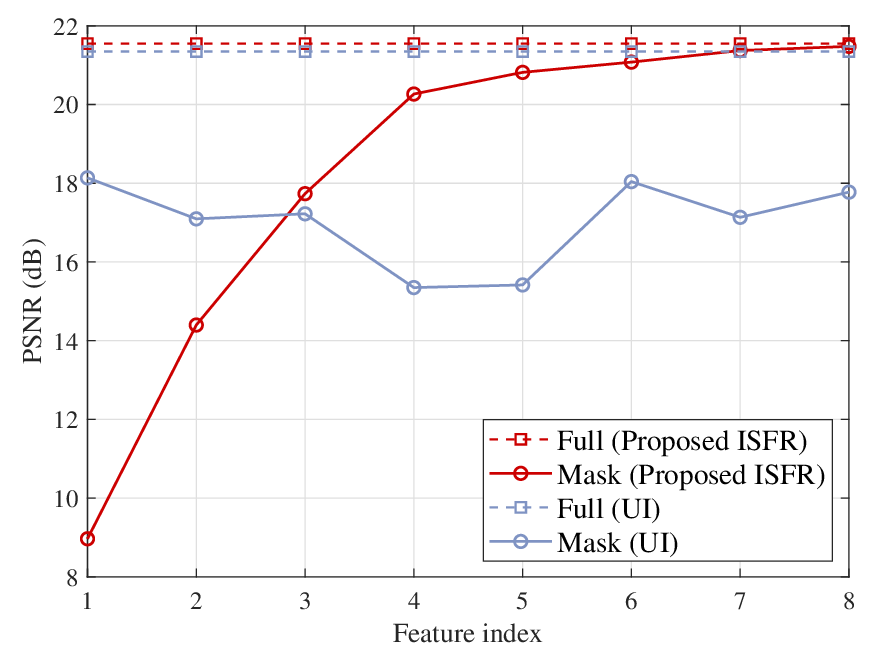}
\caption{Orderliness in ISFR via feature-wise masking.}
\label{imp-2}
\vspace{-0.1cm}
\end{figure}

\subsection{Superposition and Orderliness of Features in ISFR}
Superposition and orderliness are fundamental properties of ISFR features. Superposition implies that subsequent features progressively refine the preceding ones. As shown in Fig. \ref{imp-1}, the first feature alone captures over 80\% of the full performance, with slow gains from additional ones. Conversely, random-order ISFR and traditional UI schemes require nearly all features for high performance. Orderliness refers to the decreasing semantic contribution with increasing feature index. Fig. \ref{imp-2} illustrates that masking front-end features causes severe performance degradation, whereas masking tail-end features yields negligible loss. In contrast, UI exhibits relatively uniform sensitivity to masking across features. This confirms that ISFR establishes a structured, hierarchical semantic representation, which constitutes its core advantage over UI. These properties underpin the subsequent calculation of importance weights and an efficient SI-UEP design.

\section{Importance Evaluation and Problem Formulation}
In this section, we determine semantic importance weights via feature-wise masking and formulate an importance-weighted semantic distortion minimization problem to maximize system utility.

\subsection{Evaluation of Semantic Importance}
The feature-level semantic importance weight for SI-UEP is evaluated within the ISFR framework. By masking each individual feature in $\hat{\mathbf{Z}}_{\text{da}}$ and quantifying the performance degradation relative to the full feature baseline, the contribution of each feature is effectively characterized. Accordingly, the importance weight of the $j$-th feature $\hat{\mathbf{Z}}_{\text{da},j}$ is defined as
\begin{equation}
\hat{w}_j = g_\text{t}(h(\hat{\mathbf{Z}}_{\text{da}};\phi)) - g_\text{t}(h(\hat{\mathbf{Z}}_{\text{da}}|\hat{\mathbf{Z}}_{\text{da},j}=0;\phi)),\label{eq21}
\end{equation}
where $g_\text{t}(\cdot)$ denotes the task function. Note that these importance weights are task-dependent. In this paper, reconstruction quality is adopted as the metric for image reconstruction. However, this approach is easily extensible to other tasks, such as accuracy for image classification or mean average precision (mAP) for object detection.

The final importance weights $\mathbf{w}=[w_1,w_2,\cdots,w_N]$ are derived by normalizing the above weights, given by
\begin{equation}
\mathbf{w}=\frac{\mathbf{\hat{w}}}{\|\mathbf{\hat{w}}\|_{{\cal{L}}^1}}.\label{eq22}
\end{equation}

Furthermore, these weights remain static post model training. Therefore, they can be evaluated on the validation set during training and integrated into the model parameters as priors to guide SI-UEP. Since this approach eliminates the need for real-time task performance feedback, it is feasible at the transmitter in practical scenarios.

\subsection{SI-UEP Problem Formulation}
The transmission of $N$ semantic features extracted from the source image is considered, where each feature contains $N_\text{s}$ symbols under the same strategy. For simplicity, the feature is treated as the basic unit. Following the ISFR scheme in Section III, the features exhibit descending semantic importance, denoted by weights $w_1 > w_2 > \dots > w_N$. They are transmitted over $N$ orthogonal sub-channels with varying channel states.

The ultimate objective is to maximize task performance. However, due to the complexity of the deep neural network (DNN)-based codec, directly incorporating task performance into optimization is intractable. Therefore, this objective is transformed into minimizing the importance-weighted total semantic distortion, which consists of two components: the transmission distortion of retained features and the truncation distortion of discarded features.

For the $k$ retained features, BER serves as a proxy for transmission reliability. Assuming maximum likelihood (ML) detection over fading channels, the BER $\mu_{j}$ for the $j$-th feature is approximated as \cite{fading-BER,learned-BER1}
\begin{equation}
\begin{split}
\mu_j (p_j,m_j,\gamma_j) &\approx \frac{\sqrt{2^{m_j}} - 1}{\sqrt{2^{m_j}} \log_2 \sqrt{2^{m_j}}} \mathrm{erfc}\left( \sqrt{\frac{3 p_j \gamma_j}{2(2^{m_j} - 1)}} \right) \\
& + \frac{\sqrt{2^{m_j}} - 2}{\sqrt{2^{m_j}} \log_2 \sqrt{2^{m_j}}} \mathrm{erfc}\left( 3\sqrt{\frac{3 p_j \gamma_j}{2(2^{m_j} - 1)}} \right),\label{eq23}
\end{split}
\end{equation}
where $p_j$, $m_j$, and $\gamma_j$ denote the allocated power, modulation order, and normalized SNR, respectively, with $\mathrm{erfc}(x) = 1 - \frac{2}{\sqrt{\pi}} \int_0^x \exp(-u^2)\text{d}u$ representing the complementary error function. Conversely, each discarded feature incurs an importance-weighted penalty $D_{\text{T}}$. Therefore, the total semantic distortion is given by
\begin{equation}
J = \sum_{j=1}^k  w_j  \mu_j \bigg( p_j,m_j,\sum_{l=1}^{N} \beta_{j,l} \gamma_l \bigg) + \sum_{j=k+1}^N w_j D_{\text{T}},\label{eq24}
\end{equation}
where $\beta_{j,l}$ denotes that the $j$-th feature is allocated to the $l$-th channel.

\begin{figure*}[!t]
\centering
\includegraphics[width=16cm]{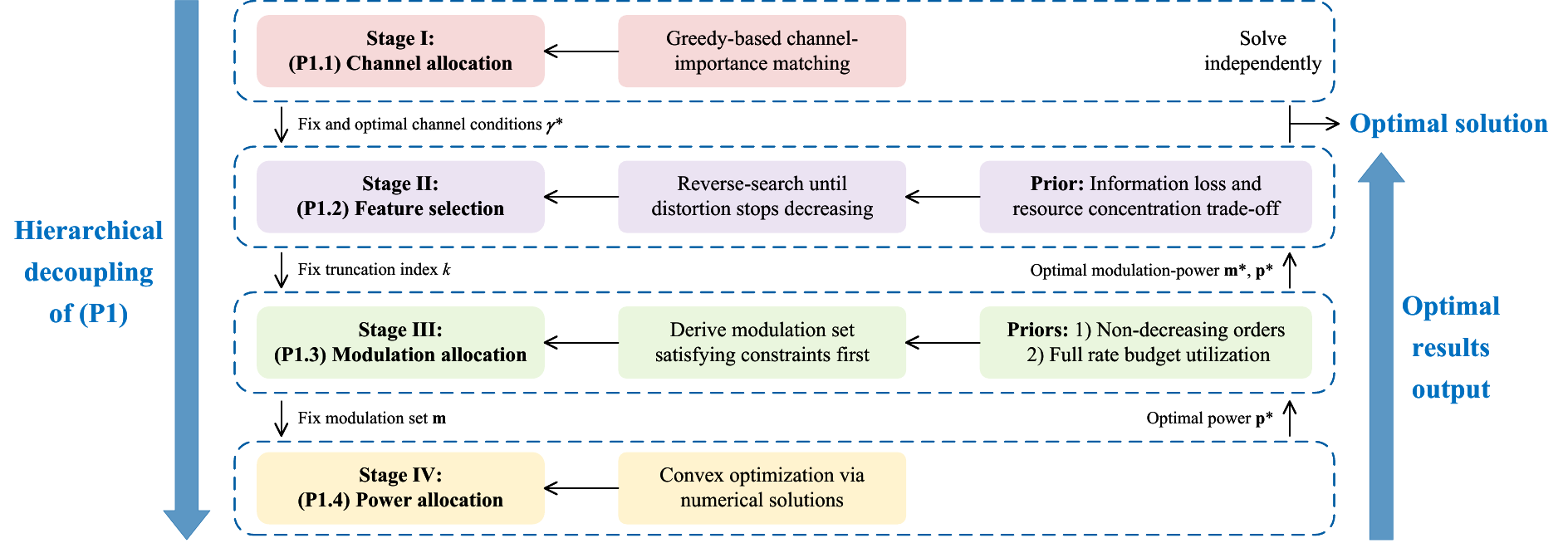}
\caption{Block diagram of the proposed OPHD algorithm for SI-UEP.}
\label{fig:algorithm}
\vspace{-0.1cm}
\end{figure*}

The total semantic distortion $J$ is minimized by jointly optimizing the truncation index $k$, channel matching indicators $\boldsymbol{\beta}$, power allocation $\mathbf{p}$, and modulation orders $\mathbf{m}$. The optimization problem is formulated as
\begin{align}\label{P1}
&(\text{P1}) \; \mathop {\min }\limits_{{{k, \boldsymbol{\beta}, \mathbf{p}}, {\mathbf{m}}}} J\\
&\qquad\quad \, \mathrm{s.t.}\;\; (\text{C1}) \; \sum\nolimits _{j = 1}^k p_j \leq NP_{\max},\notag\\
&\qquad\qquad\quad \, (\text{C2}) \; p_j \geq 0, \; \forall j,\notag\\
&\qquad\qquad\quad \, (\text{C3}) \; \frac{1}{k} \sum\nolimits _{j=1}^{k} m_j \geq M_{\text{min}} \cdot \frac{k}{N},\notag\\
&\qquad\qquad\quad \, (\text{C4}) \; m_j \in \{2, 4, 6\}, \; \forall j,\notag\\
&\qquad\qquad\quad \, (\text{C5}) \; \sum\nolimits _{j = 1}^N \beta_{j,l} = 1, \; \forall l,\notag\\
&\qquad\qquad\quad \, (\text{C6}) \; \sum\nolimits _{l = 1}^N \beta_{j,l} = 1, \; \forall j,\notag\\
&\qquad\qquad\quad \, (\text{C7}) \; \beta_{j,l} \in \left\{ {0,1} \right\}, \; \forall j,l,\notag\\
&\qquad\qquad\quad \, (\text{C8}) \; k \in \left\{ {1, 2, \cdots, N} \right\}.\notag
\end{align}
(C1) imposes the total power budget, implying that discarding features enables reallocation of power to the retained critical features. (C3) specifies the adaptive rate requirement, where reducing the feature count $k$ relaxes the rate constraints while satisfying the total delay budget. (C5)-(C7) ensure a one-to-one mapping between features and channels. 

(P1) is an intractable mixed-integer nonlinear programming (MINLP) problem due to the coupling of discrete integer variables $k, \boldsymbol{\beta}, \mathbf{m}$ and continuous variables $\mathbf{p}$. Since direct optimization is computationally prohibitive, the OPHD algorithm is introduced in the next section to address this challenge.

\section{Ordered Prior-Assisted Hierarchical Decoupling Algorithm}
In this section, the OPHD algorithm is proposed to address the challenging (P1). It is decomposed into two sub-problems: greedy-based channel matching, and joint feature selection and resource allocation utilizing pruning search and convex optimization.

\subsection{Greedy-based Channel Matching}
The first sub-problem (P1.1) aims to find the optimal channel matching matrix $\boldsymbol{\beta}$ to maximize efficiency. $N$ distinct wireless channels with CSI available at transmitter are considered, denoted as  $\boldsymbol{\gamma} = [\gamma_1, \gamma_2, \dots, \gamma_N]$. Since the optimal channel matching is independent of specific values of $k, \mathbf{p}$, and $\mathbf{m}$, a greedy strategy is employed. Accordingly, (P1.1) is expressed as
\begin{align}\label{P1.1}
&(\text{P1.1}) \; \mathop {\min }\limits_{{\boldsymbol{\beta}}} ~J\\
&\qquad\quad \, \, \mathrm{s.t.}\;\; (\text{C5}),\ (\text{C6}),\ \text{and} \ (\text{C7}).\notag
\end{align}

Specifically, features with larger semantic importance weights $w_j$ are allocated to channels with higher $\gamma_j$. This strategy reflects the principle that critical features, characterized by significant task contribution and low distortion tolerance, require enhanced protection. This mapping determines the optimal $\boldsymbol{\beta}^*$ and the sorted normalized SNR $\gamma_j^* = \sum_{l=1}^{N} \beta_{j,l}^* \gamma_l$. Consequently, the integer variable $\boldsymbol{\beta}$ is eliminated from (P1), which serves as a prerequisite for the subsequent optimization by aligning ordered features with optimal physical channels.

\subsection{Joint Feature Selection and Resource Allocation}
Given the optimal channel matching $\boldsymbol{\gamma}^*$, (P1) reduces to the joint optimization of feature selection and modulation-power allocation. This intractable MINLP problem is decomposed into three nested sub-problems via a hierarchical decoupling strategy, as shown in Fig. \ref{fig:algorithm}. Leveraging hierarchical decision dependencies, the outer layers fix decision variables to simplify the inner sub-problems. In contrast, the inner layers return optimal solutions to support the outer decisions. The formulation and solution of this top-down decoupling framework are detailed below.

\textit{1) Feature Selection via Ordered Truncation Search:} This sub-problem aims to determine the optimal feature subset for transmission. While general selection entails an exhaustive search over $2^N$ subsets, the descending order of semantic importance implies that the optimal subset corresponds to the leading $k$ features. This reduces the problem into identifying the optimal truncation index $k^*$. Assuming that the inner layer achieves optimal modulation and power allocation for any given $k$, (P1) can be simplified to (P1.2), as follows
\begin{align}\label{P1.2}
&(\text{P1.2}) \; \mathop{\min}\limits_{k} J_k = \sum_{j=1}^{k} w_j  \mu_j(p_j, m_j, \gamma_j^*) + \sum_{j=k+1}^N w_j D_{\text{T}} \\
&\qquad\quad \, \, \mathrm{s.t.}\;\; (\text{C1}),\ (\text{C2}),\ (\text{C3}),\ (\text{C4}),\ \text{and} \ (\text{C8}).\notag
\end{align}

The variation of $J_k$ with respect to $k$ reflects a fundamental trade-off between truncation loss and resource concentration gain. On the one hand, discarding tail features introduces a deterministic penalty $\sum_{j=k+1}^{N} w_j D_{\text{T}}$, representing information loss. On the other hand, reducing $k$ enables the limited resource budget to be concentrated on the remaining $k$ high-priority features, thereby yielding greater task gains.

Consequently, the optimal $k^*$ is contingent on channel conditions. In low SNR regimes, resource concentration gain dominates, causing $J_k$ to decrease as $k$ reduces. Conversely, in high SNR regimes where transmission distortion is already minimal, truncation loss becomes dominant, leading to an increase in $J_k$ as $k$ decreases. In medium SNR regimes, $J_k$ typically exhibits a unimodal profile.

Leveraging this trend, an ordered-truncation search with an early-stopping strategy is proposed. Initialized at $k=N$, the search iteratively removes the last feature and solves the inner resource allocation sub-problem. The process terminates immediately once $J_k$ begins to rise. This strategy avoids exhaustive search and ensures rapid convergence to the optimal $k^*$, as validated in Fig. \ref{UEP-4} subsequently.

\begin{algorithm}[t]
    \caption{The proposed OPHD algorithm}
    \begin{algorithmic}[1]
        \STATE \textbf{Input:} Semantic importance weights $\mathbf{w}$, normalized SNRs $\boldsymbol{\gamma}$, power budget $P_{\text{max}}$, rate budget $M_{\text{min}}$.
        \STATE \textbf{Output:} Optimal truncation index $k^*$, channel matching $\boldsymbol{\beta}^*$, modulation orders $\mathbf{m}^*$, power allocation $\mathbf{p}^*$.
        \STATE \textbf{Initialize:} $k \leftarrow N$, global minimum distortion $J_{\text{min}} \leftarrow \infty$, optimal set $\{k^*, \boldsymbol{\beta}^*,\mathbf{m}^*, \mathbf{p}^*\} \leftarrow \emptyset$.
        \STATE \textbf{Channel matching via greedy approach:} Assign $j$-th most important feature to the $j$-th highest normalized SNR channel to obtain $\boldsymbol{\beta}^*$.
        \WHILE{$k \ge 1$}
            \STATE Initialize local minimum distortion $J_{\text{local}} \leftarrow \infty$.
            \STATE \textbf{Modulation orders allocation via monotonicity-driven pruning:} Generate candidate modulation set $\mathcal{S}_{\mathbf{m}}$ satisfying (C3), (\ref{eq29}), and (\ref{eq30}).
            \FOR{each candidate vector $\mathbf{m} \in \mathcal{S}_{\mathbf{m}}$}
                \STATE \textbf{Power allocation via convex optimization:} Solve the convex problem (P1.4) using KKT conditions, Newton's method, and bisection search.
                \STATE Obtain optimal power $\mathbf{p}$ and calculate current total distortion $J_{\text{curr}}$ via (\ref{eq24}).
                
                \IF{$J_{\text{curr}} < J_{\text{local}}$}
                    \STATE Update local optima: $J_{\text{local}} \leftarrow J_{\text{curr}}$, $\mathbf{m}_{\text{local}} \leftarrow \mathbf{m}$, $\mathbf{p}_{\text{local}} \leftarrow \mathbf{p}$.
                \ENDIF
            \ENDFOR
            
            \STATE \textbf{Early stopping for ordered truncation search:}
            \IF{$J_{\text{local}} < J_{\text{min}}$}
                \STATE Update global optima: $J_{\text{min}} \leftarrow J_{\text{local}}$, $k^* \leftarrow k$, $\mathbf{m}^* \leftarrow \mathbf{m}_{\text{local}}$, $\mathbf{p}^* \leftarrow \mathbf{p}_{\text{local}}$.
            \ELSE
                \STATE Stop search as distortion $J_{\text{local}}$ starts to increase.
            \ENDIF
            
            \STATE $k \leftarrow k - 1$.
        \ENDWHILE
    \end{algorithmic}
    \label{alg:algorithm}
\end{algorithm}

\textit{2) Modulation Order Allocation via Monotonicity-Driven Pruning:} Given a fixed truncation index $k$ and the optimal inner-layer power allocation, the modulation order optimization is formulated as
\begin{align}\label{P1.3}
&(\text{P1.3}) \; \mathop{\min}\limits_{\mathbf{m}} \sum\nolimits_{j=1}^{k} w_j \mu_j(p_j, m_j, \gamma_j^*)\\
&\qquad\quad \, \, \mathrm{s.t.}\quad (\text{C1}),\ (\text{C2}),\ (\text{C3}),\ (\text{C4}).\notag
\end{align}

This remains a challenging MINLP problem due to the complexity of the approximate BER in Eq. (\ref{eq23}). However, the property that semantic importance decreases as the feature index increases is exploited. To minimize the weighted-sum BER, features with higher importance require stronger protection via lower modulation orders. Consequently, the optimal modulation orders exhibit a non-decreasing monotonicity
\begin{equation}
m_1 \le m_2 \le \dots \le m_{k}.\label{eq29}
\end{equation}

This property significantly prunes the search space for $\mathbf{m}$. Furthermore, the objective is optimized when the average modulation order approaches its lower bound, \textit{i.e.}, the rate budget is fully utilized, given by
\begin{equation}
\mathbf{m}_{\mathcal{S}} = \arg\min_{\mathbf{m}} \left| \frac{1}{k} \sum\nolimits_{j=1}^{k} m_j - M_{\text{min}} \cdot \frac{k}{N} \right|.\label{eq30}
\end{equation}

Hence, a candidate modulation set $\mathcal{S}_{\mathbf{m}}$ that satisfies Eq. (\ref{eq29}), (\ref{eq30}), and constraint (C3) is generated. Upon performing optimal power allocation in the inner layer for each candidate vector, the one yielding the minimum weighted sum BER is selected as the optimal solution for $\mathbf{m}$.

\textit{3) Power Allocation via Convex Optimization:} With fixed $k$ and $\mathbf{m}$, (P1.3) finally reduces to minimizing the weighted sum BER under total power constraint, expressed as
\begin{align}\label{P1.4}
&(\text{P1.4}) \; \mathop {\min }\limits_{{{\mathbf{p}}}} \sum_{j=1}^{k} w_j \bigg[ a_j \text{erfc} \sqrt{d_jp_j \gamma_{j}^*} + b_j \text{erfc}\bigg( c \sqrt{d_j p_j \gamma_{j}^*} \bigg) \bigg] \\
&\qquad\quad \, \, \mathrm{s.t.}\;\; (\text{C1})\ \text{and} \ (\text{C2}).\notag
\end{align}
where $a_j, b_j, c, d_j$ are constants determined by $m_j$.

Let $n(p)=\text{erfc}( u\sqrt{vp} )$, where $u,v>0$. The second derivative is strictly positive, as given by
\begin{equation}
\frac{\partial^2 n(p)}{\partial p^2} = \frac{u\sqrt{v}}{\sqrt{\pi}} e^{- u^2vp} \cdot \left( \frac{1}{2p^{3/2}} + \frac{u^2 v}{\sqrt{p}} \right) > 0.\label{eq33}
\end{equation}
Since the objective function of (P1.4) is a non-negative linear combination of $n(p)$, and the constraints are linear, (P1.4) is a convex optimization problem. 

According to the Karush-Kuhn-Tucker (KKT) conditions, the optimal solution $\mathbf{p}^*=[p_1^*,p_2^*,\cdots,p_{k^*}^*]$ satisfies
\begin{equation}
\begin{cases}
t(p_{j}^*)=w_j \sqrt{\frac{d_j\gamma_{j}^*}{\pi p_{j}^*}} \left[ a_j e^{-d_j\gamma_{j}^* p_{j}^*} + b_j c e^{-c^2 d_j\gamma_{j}^* p_{j}^*} \right] = \lambda^* \\
\sum_{j=1}^{k} p_{j}^* = NP_{\text{max}}
\end{cases}
\end{equation}
where $\lambda^*$ is the Lagrange multiplier. Since $t(p_j)$ is monotonically decreasing with respect to $p_j$, it can be solved numerically using Newton's method for a given $\lambda$. Subsequently, the optimal $\lambda^*$ is determined via a bisection search to satisfy the power budget.

\begin{figure}[!t]
\centering
\includegraphics[width=6cm]{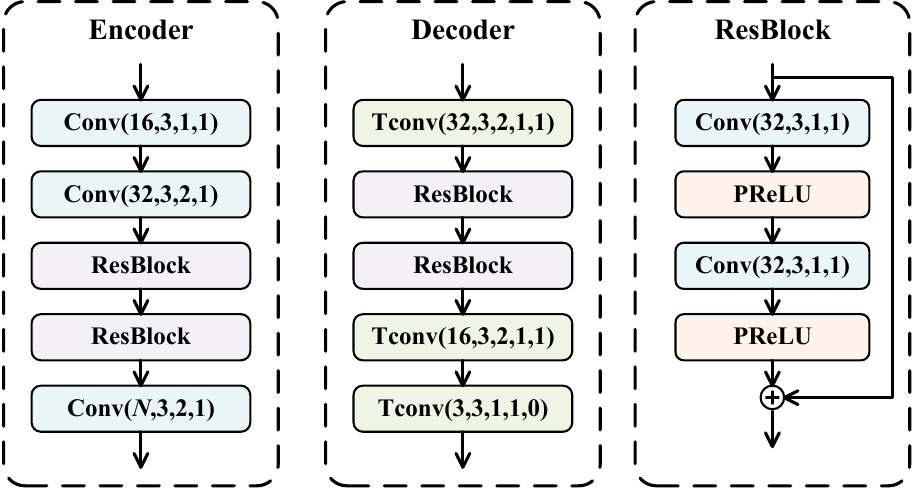}
\caption{The model architecture of semantic codec. Conv($c$,$k$,$s$,$p$) denotes a convolutional layer with $c$ output channels, kernel size $k$, stride $s$, and padding $p$. Tconv($c$,$k$,$s$,$p$,$p_\text{o}$) denotes a transposed convolutional layer with $c$ output channels, kernel size $k$, stride $s$, padding $p$, and output padding $p_\text{o}$.}
\label{CNN_model}
\vspace{-0.1cm}
\end{figure}

\subsection{Complexity Analysis}
The proposed OPHD algorithm, summarized in \textbf{Algorithm 1}, decomposes the intractable (P1) into four relatively independent solution procedures. Here, the ordered importance plays a crucial role in significantly reducing the computational complexity. Specifically, channel matching holds a complexity of $\mathcal{O}(N \log N)$. Leveraging the monotonicity in Eq. (\ref{eq29}) and the full rate budget utilization in Eq. (\ref{eq30}), the complexity of modulation order search is reduced to $\mathcal{O}(k^{N_\text{m}-2})$, where $N_\text{m}$ is the number of modulation orders. For power allocation, the complexity is $\mathcal{O}(I_\text{b}I_\text{n}k)$, where $I_\text{b}$ and $I_\text{n}$ denote the iteration counts for the bisection search and Newton's method, respectively. Since $I_\text{b}$ and $I_\text{n}$ are constants independent of $N$, this simplifies to $\mathcal{O}(k)$. Given that the ordered truncation search iterates at most $N$ times, the overall complexity of OPHD is $\mathcal{O}(N \log N) + \sum\nolimits_{k=k_\text{min}}^{N} \mathcal{O}(k^{N_\text{m}-2}) \cdot \mathcal{O}(k) \approx \mathcal{O}(N^{N_\text{m}})$. For comparison, the exhaustive search scheme without ordered priors requires $\mathcal{O}(N(N_\text{m}+1)^N)$. In summary, the proposed OPHD algorithm successfully reduces the complexity of solving (P1) from exponential to polynomial order.

\begin{figure}[!t]
\setlength{\abovecaptionskip}{-0.05cm}   
\centering
\includegraphics[width=6.5cm]{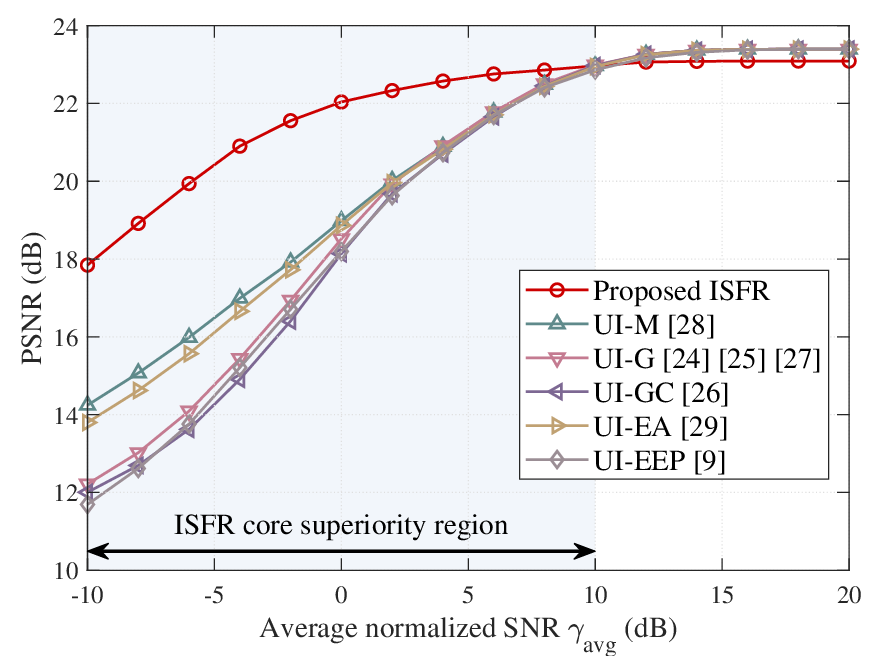}
\caption{Impact of importance distributions on PSNR with different $\gamma_{\text{avg}}$, $P_{\text{max}}=2~\text{W}$, and $M_{\text{min}}=4$ on CIFAR-10.}
\label{PSNR}
\vspace{-0.2cm}
\end{figure}

\section{Simulation Results}
In this section, we use simulations to verify the effectiveness of the proposed ISFR scheme and the OPHD algorithm.
\subsection{Simulation Settings}
\textit{1) Datasets and Metrics:} For low-resolution images, the CIFAR-10 dataset is utilized, partitioned into 40,000 training, 10,000 validation, and 10,000 testing samples. For high-resolution images, the ImageNet validation set is adopted for training following \cite{DJSCC-V}. It is split into 40,000 training and 10,000 validation samples, and center-cropped to $256 \times 256$. For testing, 100 2k-resolution images from the DIV2K dataset are employed.
To evaluate reconstruction quality, three key metrics are employed: peak signal-to-noise ratio (PSNR), structural similarity index measure (SSIM), and learned perceptual image patch similarity (LPIPS). These metrics quantify pixel, structural, and perceptual fidelity, respectively.

\textit{2) Parameter Settings:} The semantic codec utilizes CNN-based architecture, as detailed in Fig. \ref{CNN_model}. Features are partitioned along the channel dimension to capture distinct semantic contributions. Unless otherwise specified, the number of features is set to $N=8$, with quantization bits $N_\text{b}=2$. During training, the random BER upper bound $\text{BER}_\text{max}=0.2$, and the temperature coefficient $T = \min\{3 \cdot \text{epoch} + 1, 300\}$. For both training stages, the model is trained for 100 epochs with a learning rate of $10^{-3}$, followed by another 100 epochs at $10^{-4}$, using a batch size of 64. Semantic importance weights are derived from PSNR variations on the validation set.
During inference, the average power budget $P_\text{max}$ ranges from 0.4 W to 4 W, and the average rate budget $M_\text{min}$ varies from 2 to 6. To simulate channel heterogeneity across features (\textit{e.g.}, time-varying or frequency-selective fading),  $N$ distinct SNR values are randomly sampled within a $\pm5$ dB range around the average normalized SNR $\gamma_{\text{avg}}$. Additionally, $T$ is fixed at 300, the feature discarding penalty $D_\text{T}=0.22$, and the convergence threshold for the numerical solution is $10^{-5}$.

\begin{figure}[!t]
\setlength{\abovecaptionskip}{-0.05cm}   
\centering
\includegraphics[width=6.5cm]{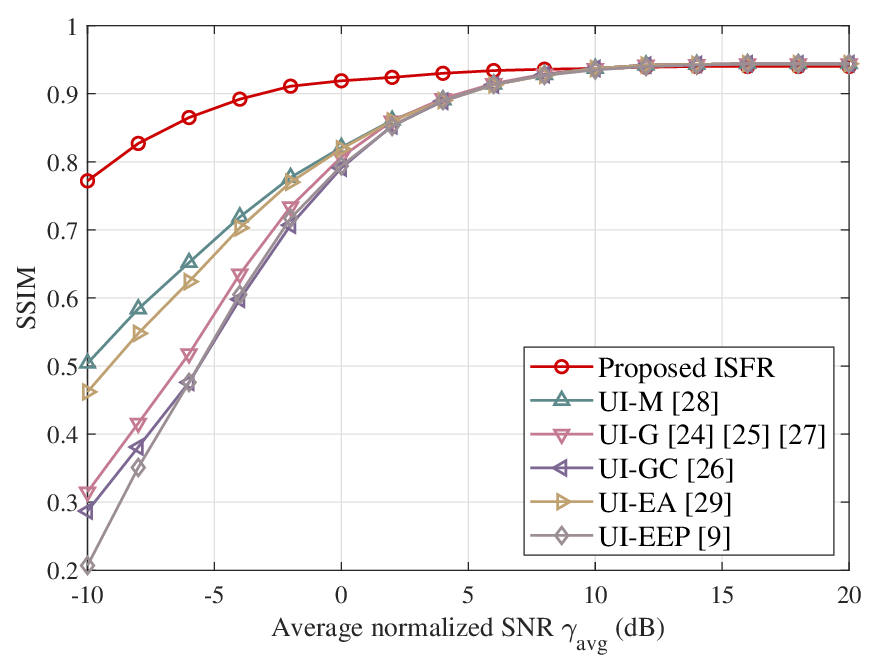}
\caption{Impact of importance distributions on SSIM with different $\gamma_{\text{avg}}$, $P_{\text{max}}=2~\text{W}$, and $M_{\text{min}}=4$ on CIFAR-10.}
\label{SSIM}
\vspace{-0.2cm}
\end{figure}

\begin{figure}[!t]
\setlength{\abovecaptionskip}{-0.05cm}   
\centering
\includegraphics[width=6.5cm]{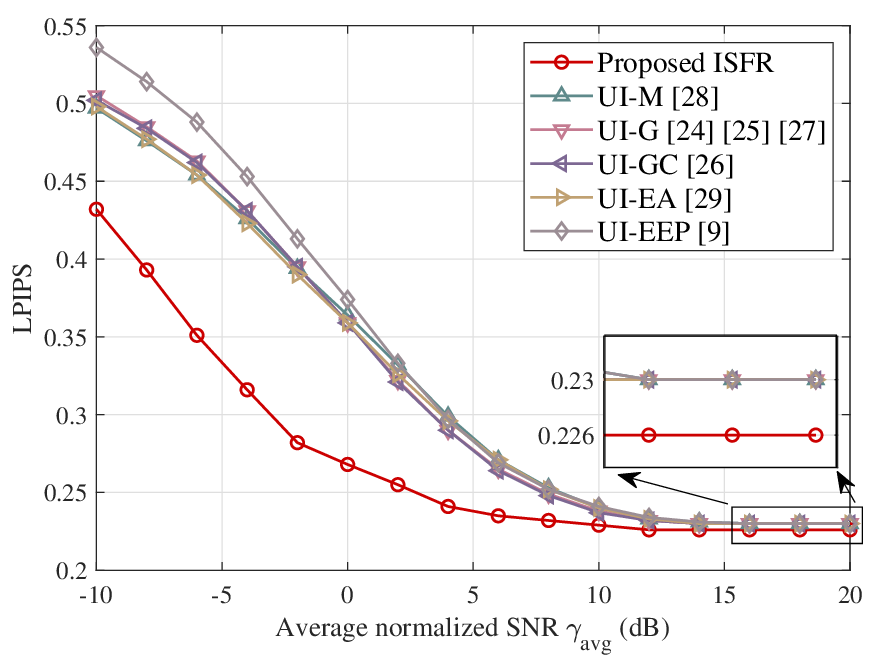}
\caption{Impact of importance distributions on LPIPS with different $\gamma_{\text{avg}}$, $P_{\text{max}}=2~\text{W}$, and $M_{\text{min}}=4$ on CIFAR-10.}
\label{LPIPS}
\vspace{-0.2cm}
\end{figure}

\textit{3) Benchmarks:} To validate the advantages of the proposed ISFR scheme over traditional UI paradigms, five baselines are introduced for comparison:
\begin{itemize}
\item \textbf{UI with Mask-based Evaluation (UI-M)\cite{masking}}: This scheme quantifies importance by measuring the performance degradation caused by feature-wise masking relative to the full feature baseline. It aligns with the strategy in ISFR and represents a direct and objective method for feature importance evaluation.
\item \textbf{UI with Gradient Analysis (UI-G)\cite{grad1,grad2,grad3}}: This scheme derives importance weights from the average gradient of the loss function with respect to feature values. It effectively captures feature contributions to the task and is widely adopted in task-oriented scenarios.
\item \textbf{UI with Gradient-Correlation Analysis (UI-GC)\cite{grad-corr}}: Building upon UI-G, this scheme incorporates inter-feature correlations into importance weights.
\item \textbf{UI with Error Approximation (UI-EA)\cite{grad-act}}: This scheme approximates UI-M using the squared product of gradients and activation values via first-order Taylor expansion, thereby reducing computational complexity.
\item \textbf{UI with Equal Error Protection (UI-EEP)\cite{DJSCC}}: This scheme relies solely on end-to-end training for channel adaptation, without explicit feature importance analysis or SI-UEP mechanisms. 
\end{itemize}

\begin{figure*}[t]
\centering
\subfloat[]{
\label{bar-a}
\includegraphics[width=5.9cm]{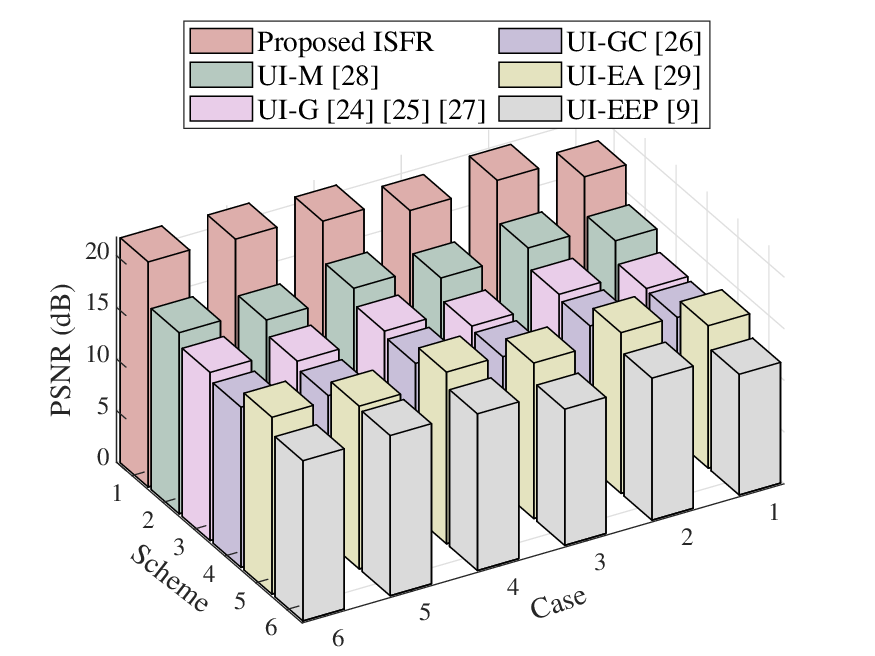}}
\subfloat[]{
\label{bar-b}
\includegraphics[width=5.9cm]{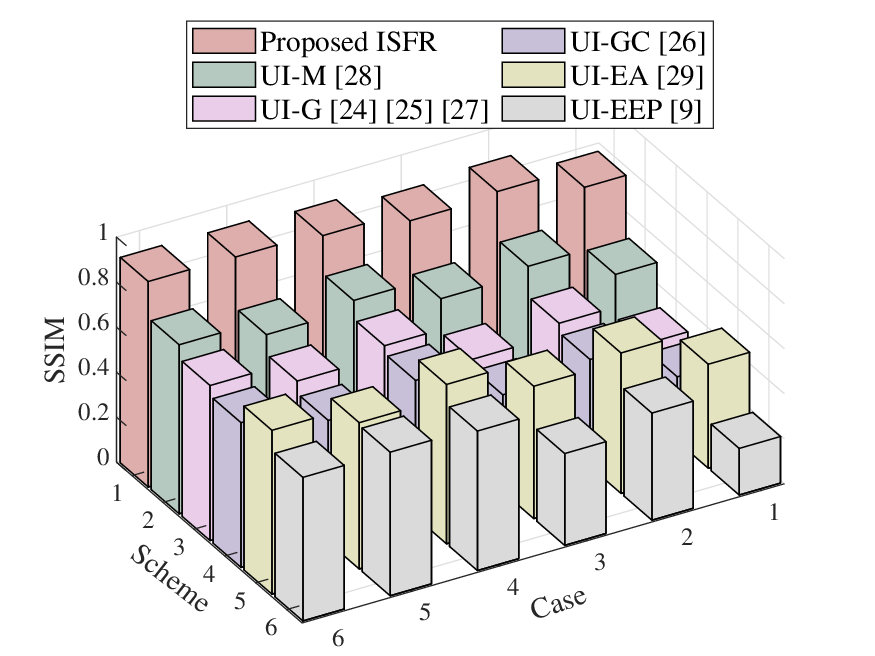}}
\subfloat[]{
\label{bar-c}
\includegraphics[width=5.9cm]{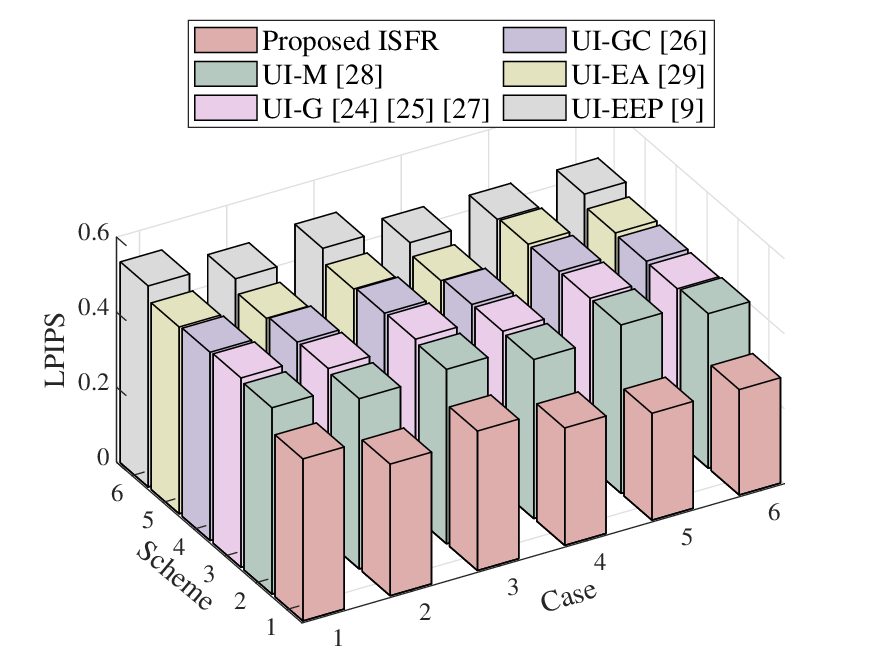}}
\caption{(a) PSNR, (b) SSIM, and (c) LPIPS performance under adverse conditions on CIFAR-10.}
\label{bar}
\vspace{-0.2cm}
\end{figure*}

\subsection{Impact of Importance Distributions on Performance}
Fig. \ref{PSNR} illustrates the PSNR performance of different importance distributions across various average normalized SNR $\gamma_{\text{avg}}$. In low-to-medium SNR region (\textit{i.e.}, $\gamma_{\text{avg}} < 10 \ \text{dB}$), the proposed ISFR-SI-UEP scheme significantly outperforms all five baselines. Specifically, at $\gamma_{\text{avg}} = -6 \ \text{dB}$, it achieves performance gains of 24.7\%, 41.5\%, 46.4\%, 28.1\%, and 44.9\%, respectively. This superiority stems from the concentrated information distribution in ISFR, which positions high-priority semantic features at leading indices. Consequently, the SI-UEP strategy effectively protects these critical features against poor channel conditions, ensuring essential semantic reconstruction. In contrast, the traditional UI scheme adopts uniform information distribution. As a result, channel degradation easily corrupts the scattered key information, making targeted protection infeasible and leading to a rapid performance drop.
Conversely, at high $\gamma_{\text{avg}}$ (\textit{i.e.}, $\gamma_{\text{avg}} > 10 \ \text{dB}$), performance saturates across all schemes. ISFR-SI-UEP performs slightly lower than the top UI baseline by 1.3\%. This indicates that the core advantage of ISFR lies in low-to-medium $\gamma_{\text{avg}}$ regions. It prioritizes superior robustness under harsh channels, at the cost of a marginal loss in peak information capacity under ideal conditions.

Further, Fig. \ref{SSIM} and Fig. \ref{LPIPS} illustrate the SSIM and LPIPS performance. ISFR-SI-UEP maintains significant superiority at low $\gamma_{\text{avg}}$. Particularly, at $\gamma_{\text{avg}} = -6 \ \text{dB}$, SSIM is improved by over 32.7\% and LPIPS is reduced by over 22.7\%. At high $\gamma_{\text{avg}}$, SSIM of ISFR is comparable to UI, while LPIPS is slightly superior, differing from the trend in PSNR. 
These results indicate that ISFR-SI-UEP prioritizes structural and perceptual semantic preservation despite a minor loss of peak capacity. The PSNR loss mostly corresponds to non-essential details with negligible impact on both human and machine vision. However, this capacity constraint may vary with data complexity, which is further examined in the subsequent DIV2K tests.

\subsection{Advantages under Adverse Conditions}
To validate the robustness advantages of the proposed ISFR scheme under adverse conditions, six test cases are configured with low average normalized SNR $\gamma_{\text{avg}}$, low power budget $P_{\text{max}}$, and high rate requirement $M_{\text{min}}$:
\begin{itemize}
\item \textbf{Case 1:} $\gamma_{\text{avg}} = -10 \ \text{dB}$, $P_{\text{max}} = 2 \ \text{W}$, $M_{\text{min}} = 4$.
\item \textbf{Case 2:} $\gamma_{\text{avg}} = -6 \ \text{dB}$, $P_{\text{max}} = 2 \ \text{W}$, $M_{\text{min}} = 4$.
\item \textbf{Case 3:} $\gamma_{\text{avg}} = 0 \ \text{dB}$, $P_{\text{max}} = 0.4 \ \text{W}$, $M_{\text{min}} = 4$.
\item \textbf{Case 4:} $\gamma_{\text{avg}} = 0 \ \text{dB}$, $P_{\text{max}} = 0.8 \ \text{W}$, $M_{\text{min}} = 4$.
\item \textbf{Case 5:} $\gamma_{\text{avg}} = 0 \ \text{dB}$, $P_{\text{max}} = 2 \ \text{W}$, $M_{\text{min}} = 6$.
\item \textbf{Case 6:} $\gamma_{\text{avg}} = 0 \ \text{dB}$, $P_{\text{max}} = 2 \ \text{W}$, $M_{\text{min}} = 5$.
\end{itemize}

Fig. \ref{bar} compares the performance in terms of PSNR, SSIM, and LPIPS. As shown, ISFR-SI-UEP consistently outperforms all five baselines in all scenarios. For instance, it achieves PSNR gains of over 25\% at low SNR (Case 1-2), over 23\% at low power (Case 3-4), and over 24\% at high rate (Case 5-6), compared to the best performing baseline. Similar trends are observed in SSIM and LPIPS, demonstrating the structural and perceptual consistency of ISFR. These results underscore the superior robustness of ISFR-SI-UEP over traditional UI schemes under adverse transmission conditions. In such environments, resources are insufficient to fully protect all extracted features, resulting in severe performance degradation in the UI. In contrast, ISFR concentrates resources on the most semantically significant features, ensuring essential fidelity. This further reinforces SemCom's inherent advantage: reliable transmission even in extremely harsh environments.

\begin{figure}[!t]
\setlength{\abovecaptionskip}{-0.05cm}   
\centering
\includegraphics[width=6.5cm]{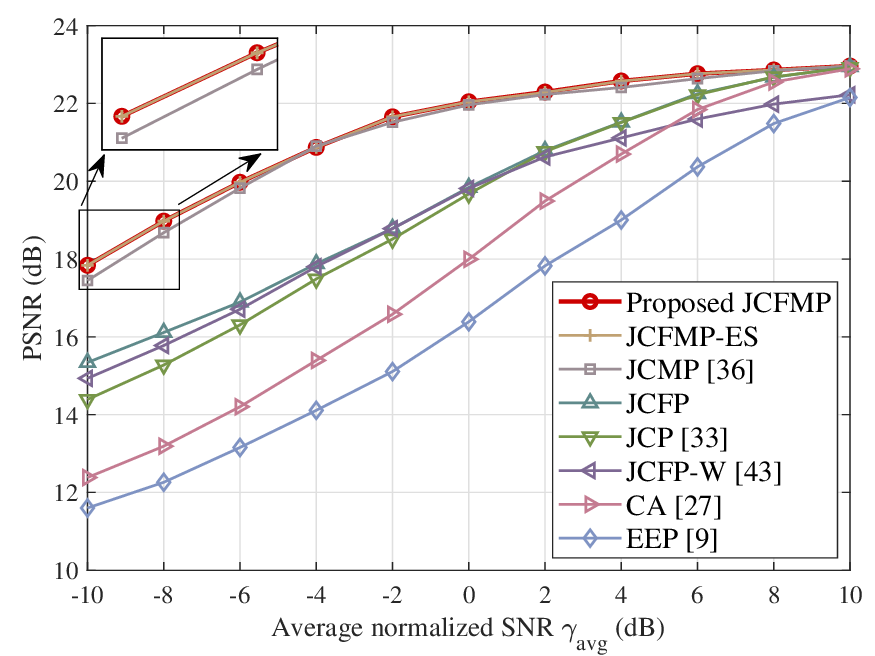}
\caption{Impact of SI-UEP strategies on PSNR within ISFR, with different $\gamma_{\text{avg}}$, $P_{\text{max}}=2~\text{W}$ and $M_{\text{min}}=4$ on CIFAR-10.}
\label{UEP-4}
\vspace{-0.2cm}
\end{figure}

\subsection{Impact of SI-UEP strategies on Performance}
Additionally, Fig. \ref{UEP-4} evaluates the impact of SI-UEP strategies within ISFR. Comparison includes the OPHD-based joint channel, feature, modulation, and power allocation (JCFMP), joint channel, feature, and power allocation (JCFP), channel-only allocation (CA) \cite{grad3}, and EEP. Furthermore, JCFMP-ES is included to represent JCFMP with an exhaustive search for feature selection. Baselines without feature selection are denoted as JCMP\cite{learned-BER1} and JCP\cite{com4cv}, while JCFP-W represents the classical waterfilling-based power allocation for sum-rate maximization \cite{water}. Note that unspecified “SI-UEP” defaults to JCFMP for ISFR and JCMP for UI, as the latter excludes feature discarding during training.

\begin{figure}[!t]
\centering
\subfloat[]{
\label{vis-a}
\includegraphics[width=6.2cm]{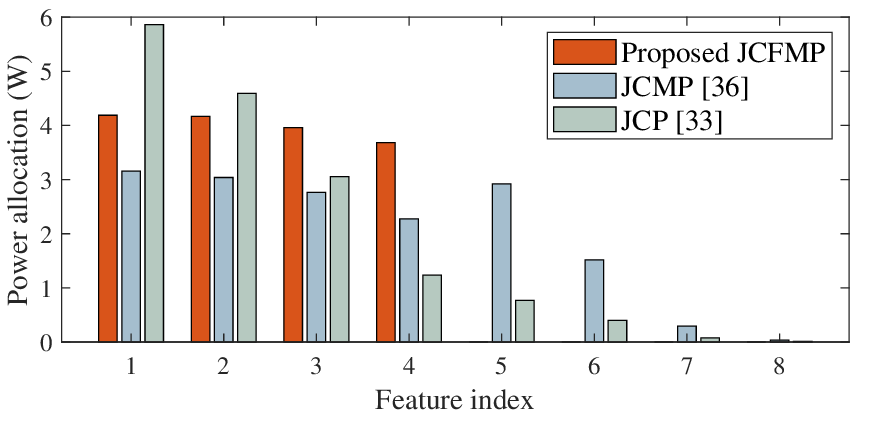}}\vspace{-0.05cm} 
\subfloat[]{
\label{vis-b}
\includegraphics[width=6.2cm]{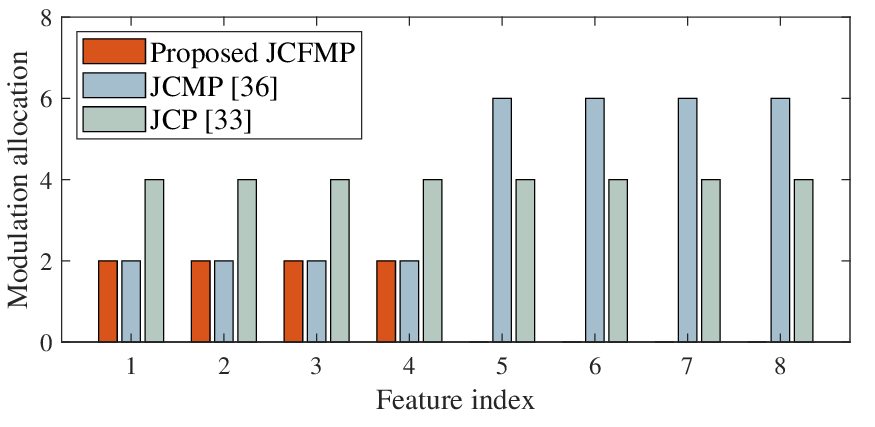}}\vspace{-0.05cm} 
\subfloat[]{
\label{vis-c}
\includegraphics[width=6.4cm]{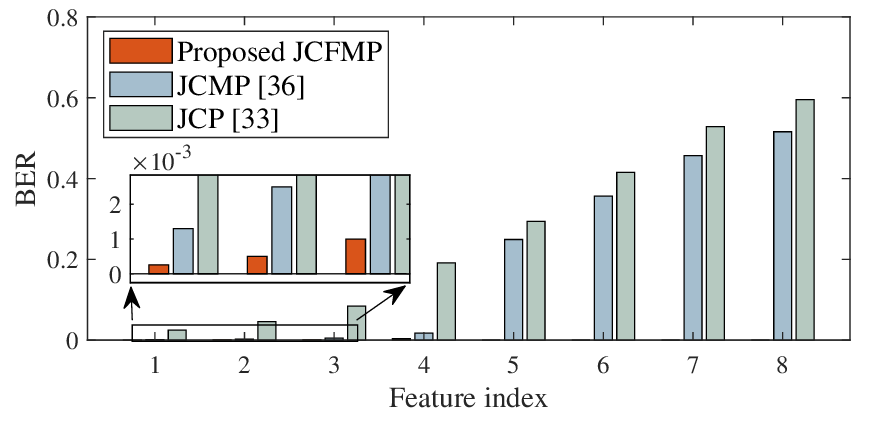}}
\caption{Resource allocation and BER performance visualization under different SI-UEP strategies.}
\label{vis}
\vspace{-0.1cm}
\end{figure}

The proposed JCFMP scheme achieves optimal performance, validating the necessity of joint multi-dimensional resource optimization. Its performance aligns with the exhaustive search-based JCFMP-ES, confirming that the early stopping strategy effectively reduces computational complexity with negligible performance loss. Moreover, modulation optimization yields over 15\% gain at low $\gamma_{\text{avg}}$, underscoring its critical role.
In contrast, feature selection yields negligible gains for JCMP. Since 4QAM (\textit{i.e.}, the most robust format) is already uniformly adopted at low $\gamma_{\text{avg}}$, retaining fewer features does not further relax the rate constraint. Consequently, the benefits translate to improved spectral efficiency and reduced latency. Only 41\% of total subcarriers are required at $\gamma_{\text{avg}} = -10 \ \text{dB}$.
Furthermore, the classical waterfilling scheme exhibits a noticeable gap compared to JCFMP. This stems from its rate-centric objective, which fails to capture semantic distortion, and the exclusion of modulation orders from optimization.

To further present the distinct strategies, Fig. \ref{vis} visualizes the resource allocation under JCFMP, JCMP, and JCP. JCFMP achieves optimal BER performance by concentrating resources on the first four reserved features. JCMP lacks feature selection and still allocates power to less important features, resulting in slightly inferior performance compared to JCFMP. Meanwhile, although JCP prioritizes front-end features in power allocation, it neglects modulation orders that dominate BER, leading to significantly degraded performance. These results validate the effectiveness of the proposed JCFMP scheme for SI-UEP.

\subsection{Feature Restructuring Effect}
\begin{figure}[!t]
\setlength{\abovecaptionskip}{-0.05cm}   
\centering
\includegraphics[width=6.5cm]{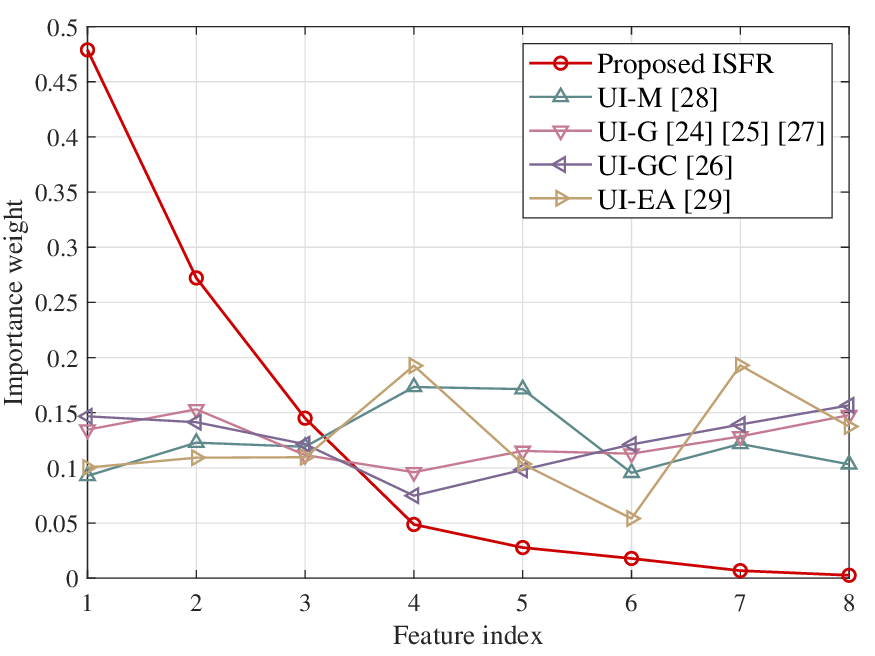}
\caption{Semantic importance weights under different feature distributions and evaluation schemes on CIFAR-10.}
\label{imp-3}
\vspace{-0.1cm}
\end{figure}

\begin{figure}[!t]
\centering
\subfloat[]{
\label{imp-4-a}
\includegraphics[width=6.3cm]{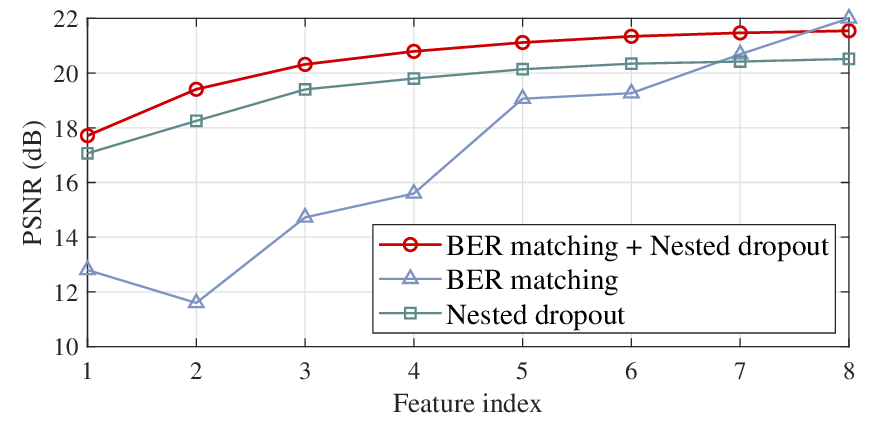}}
\vspace{-0.05cm} 
\subfloat[]{
\label{imp-4-b}
\includegraphics[width=6.3cm]{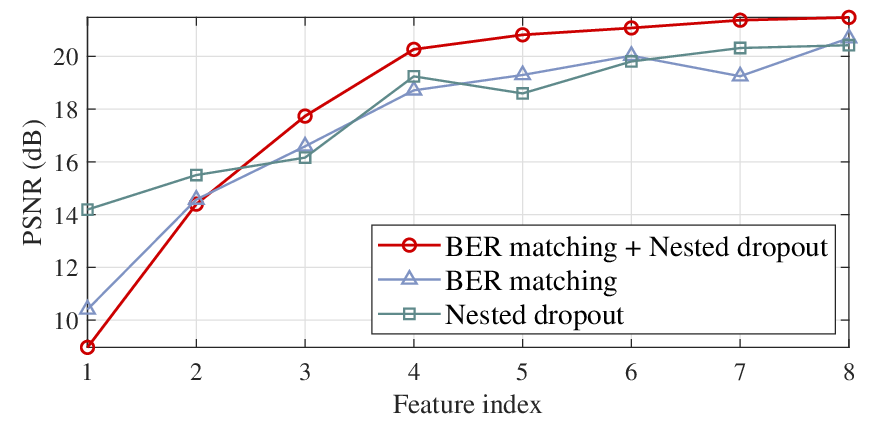}}
\caption{Impact of BER matching and nested dropout modules with (a) feature accumulation, (b) feature-wise masking on CIFAR-10.}
\label{imp-4}
\vspace{-0.1cm}
\end{figure}

The semantic importance weights of the proposed ISFR and four UI baselines are shown in Fig. \ref{imp-3}. ISFR exhibits a distinctly different importance-decaying distribution, whereas all baselines derive roughly uniform weights with minor fluctuations regardless of evaluation methods. This stems from the inherent limitation of post-training evaluation in UI schemes. Without extra constraints, these models tend to learn uniform distributions, which hinders efficient SI-UEP.

Subsequently, the benefits of jointly adopting the BER matching and nested dropout modules are explored. Fig. \ref{imp-4} presents results from feature accumulation and feature-wise masking. The joint approach exhibits higher stability, superior performance, and more pronounced importance differentiation compared to individual modules. This validates the effectiveness of integrating these modules within ISFR framework.

\begin{figure*}[t]
\centering
\subfloat[]{
\label{div2k-a}
\includegraphics[width=5.9cm]{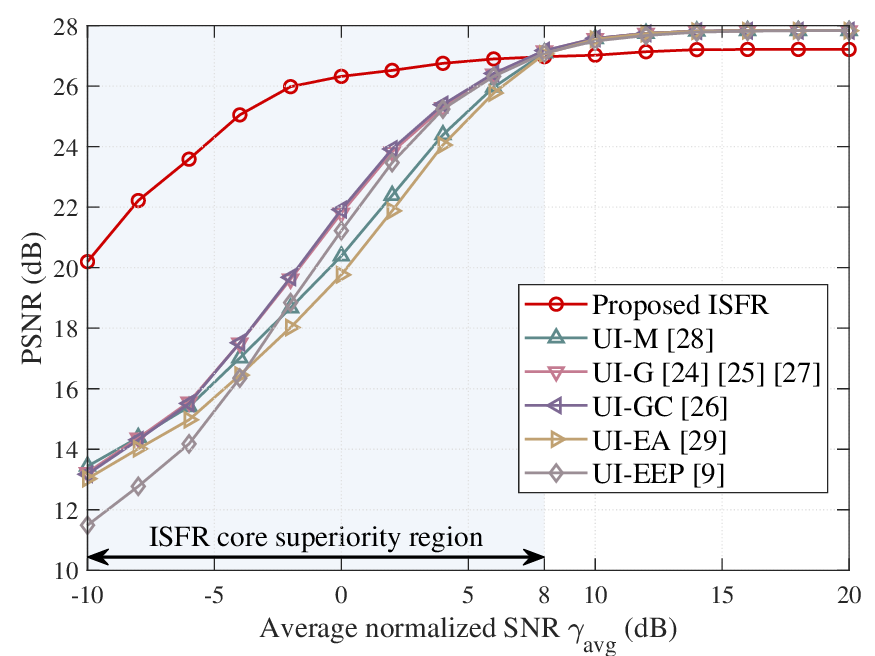}}
\subfloat[]{
\label{div2k-b}
\includegraphics[width=5.9cm]{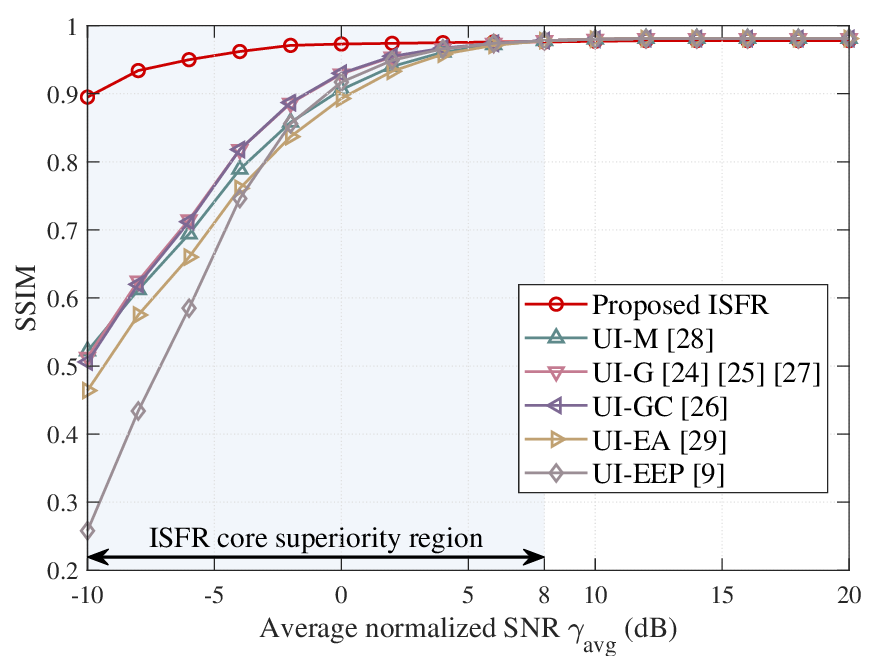}}
\subfloat[]{
\label{div2k-c}
\includegraphics[width=5.9cm]{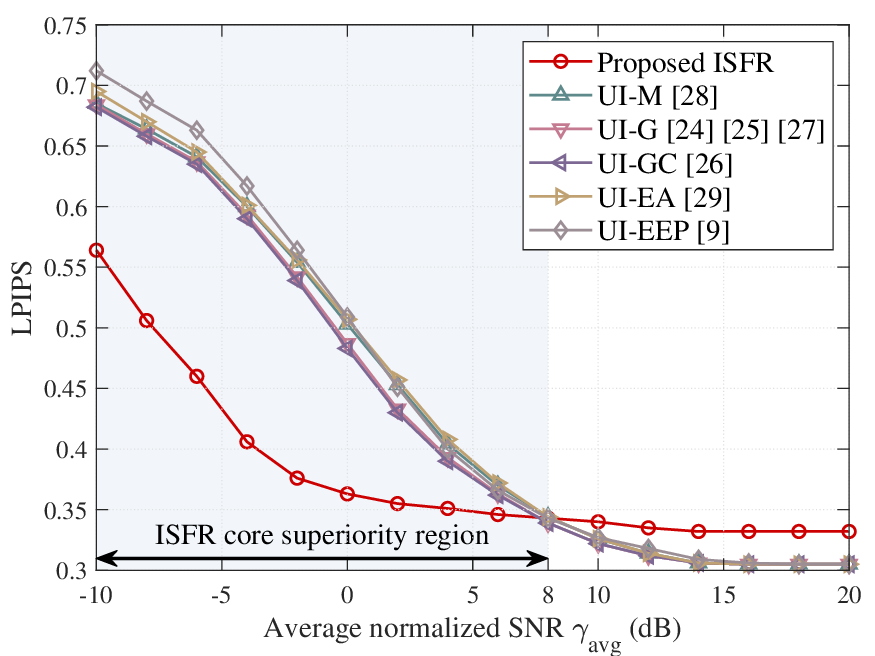}}
\caption{Impact of importance distributions on (a) PSNR, (b) SSIM, and (c) LPIPS with different $\gamma_{\text{avg}}$, $P_{\text{max}}=2~\text{W}$, and $M_{\text{min}}=4$ on DIV2K.}
\label{div2k}
\vspace{-0.1cm}
\end{figure*}

\begin{figure*}[!t]
\setlength{\abovecaptionskip}{-0.05cm}   
\centering
\includegraphics[width=17.5cm]{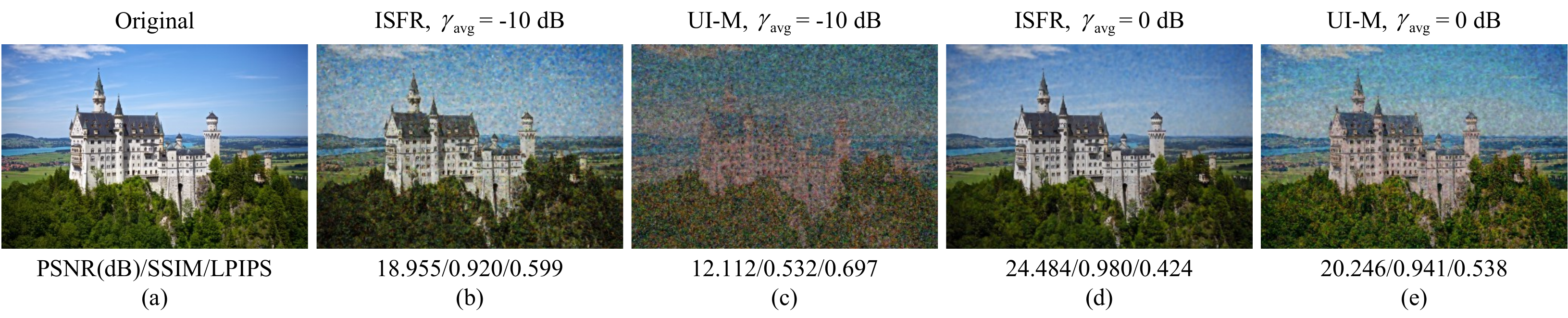}
\caption{Visualization results of the reconstructed images on DIV2K.}
\label{vis_div2k}
\vspace{-0.1cm}
\end{figure*}

\subsection{Generalization Verification on DIV2K Dataset}
To verify the generalization of the proposed ISFR scheme, its performance is further evaluated on the high-resolution DIV2K dataset. Fig. \ref{div2k} presents the PSNR, SSIM, and LPIPS results.
Consistent with the findings on CIFAR-10, ISFR-SI-UEP achieves over 20\% robustness gains at low $\gamma_{\text{avg}}$ across all metrics. Visualization results in Fig. \ref{vis_div2k} confirm that ISFR yields clearer reconstructed images than UI under degraded channel conditions. This outcome validates the effectiveness of ISFR for SI-UEP across diverse data distributions. 
However, a slight performance drop relative to UI occurs at high $\gamma_{\text{avg}}$ across all metrics. Unlike on CIFAR-10, the LPIPS advantage diminishes due to the richer high-frequency details on DIV2K. In this context, ISFR acts as a semantic filter, prioritizing global structures over noise-sensitive fine-grained textures, leading to marginal perceptual loss under ideal conditions.

This highlights a fundamental trade-off between structured robustness and unconstrained capacity. While UI approximates unconstrained optimization to maximize capacity, ISFR imposes an importance-ordered prior for structured hierarchical semantic representation. Although this constraint leads to a slight performance loss, it is essential for enabling efficient, robust, and low-complexity SI-UEP. Crucially, for SemCom, maintaining essential semantic fidelity under extremely adverse conditions is prioritized over maximizing peak fidelity in ideal regimes. Consequently, this resilience-centric architecture delivers substantial utility for 6G edge intelligence scenarios involving harsh channels and limited resources.

\section{Conclusion}
In this paper, the ISFR scheme and SI-UEP have been investigated to address the insufficient utilization of importance distribution in existing schemes.
Specifically, the ISFR scheme has been proposed to concentrate critical semantics into front-end features by employing unequal retention probabilities and distortion levels.
For SI-UEP, a joint optimization problem that jointly optimizes channel matching, feature selection, modulation schemes, and power allocation has been formulated to minimize total semantic distortion.
To efficiently solve this complex problem, a low-complexity OPHD algorithm has been developed that integrates a greedy strategy, search space pruning, and convex optimization.
Simulation results demonstrate that ISFR achieves performance gains exceeding 23\% over traditional UI schemes \cite{grad1,grad2,grad3,grad-corr,masking,grad-act} under adverse conditions with low SNR, limited power, and high data rates. 
This superior robustness is attributed to pronounced importance differentiation spanning two orders of magnitude. 
Under ideal conditions, ISFR incurs only a marginal peak loss of 1.3\%, highlighting the inherent trade-off between robustness and informativeness.
These findings confirm that the resilience-centric design of ISFR is highly beneficial for harsh channel conditions and resource-constrained scenarios that require effective SI-UEP.
Future work will explore advanced training strategies to control dynamically the importance distribution based on real-time CSI, further improving adaptability to varying transmission conditions.


 




\vfill

\end{document}